\documentclass[12pt]{article}
\pdfoutput=1

\usepackage{epsfig}
\usepackage{psfrag}
\usepackage{latexsym}
\usepackage{amsmath}
\usepackage{amssymb}
\usepackage{amsfonts}
\usepackage{bbold}
\usepackage{graphicx}
\usepackage{cite}
\usepackage{pstricks}
\usepackage{inputenc}
\usepackage{verbatim}
\usepackage{slashed}
\usepackage{array}
\newpsobject{grilla}{psgrid}{subgriddiv=1,griddots=10,gridlabels=6pt}
\usepackage{hyperref}

\def\eq#1{{eq.~(\ref{#1})}}

\def\sec#1{{sec.~(\ref{#1})}}

\def\vev#1{\left\langle #1\right\rangle}

\def\Im{\mbox{Im}\,}
\def\Re{\mbox{Re}\,}

\def\di{\mbox{d}}

\def\hbar{\hspace{0pt}\raisebox{1pt}{$-$} \hspace{-7pt} h}

\newcommand{\be}{\begin{equation}}
\newcommand{\ee}{\end{equation}}
\newcommand{\bea}{\begin{eqnarray}}
\newcommand{\eea}{\end{eqnarray}}
\newcommand{\nn}{\nonumber}

\def\cU{{\cal U}}

\def\5{\overline 5}

\renewcommand{\Re}{\mbox{Re}}
\renewcommand{\Im}{\mbox{Im}}

\def\vev#1{\left\langle #1\right\rangle}
\let\vev\vev

\def\roughly#1{\mathrel{\raise.3ex\hbox{$#1$\kern-.75em
      \lower1ex\hbox{$\sim$}}}} 

\def\e6{E(6)}
\def\321{$SU(3)_{c}\otimes SU(2)_L \otimes U(1)$}
\def\10{SO(10)}

\def\422{SU(4) $\otimes$ SU(2) $\otimes$ SU(2)}

\newcommand{\beq}{\begin{equation}}
\newcommand{\eeq}{\end{equation}}
\newcommand{\bac}{\beq\begin{array}}
\newcommand{\eac}{\end{array}\eeq}
\newcommand{\ba}{\begin{array}}
\newcommand{\ea}{\end{array}}
\newcommand{\esp}{\end{split}}
\newcommand{\bsp}{\begin{split}}

\newcommand{\bmat}{\begin{pmatrix}}
\newcommand{\emat}{\end{pmatrix}}
\def\author#1{\begin{center} #1\end{center}}

\newcommand{\appendixA}{\setcounter{equation}{0}
\def\theequation{\rm{A}.\arabic{equation}}\section*}

\begin{document}
\begin{titlepage}
\vspace*{-1cm}
\phantom{hep-ph/***}

\hfil{NIKHEF 2010-046}
\hfill{TUM-HEP-781/10}
\hfill{DFPD/2010/TH/19}

\vskip 1.3cm
\begin{center}
\mathversion{bold}
{\Large \bf  Constraining Flavour Symmetries At The EW Scale I:\\The {$A_4$} Higgs Potential}
\mathversion{normal}
\end{center}

\vskip 0.5  cm

\begin{center}
 {\large Reinier de Adelhart Toorop}~$^{a)}$\footnote{e-mail address: reinier.de.adelhart.toorop@nikhef.nl},
{\large Federica Bazzocchi}~$^{b)}$\footnote{e-mail address: fbazzoc@few.vu.nl},
 \vskip 0.1cm
{\large Luca Merlo}~$^{c)}$\footnote{e-mail address: luca.merlo@ph.tum.de}
and {\large Alessio Paris}~$^{d)}$\footnote{e-mail address: alessio.paris@pd.infn.it}
\\
\vskip .1cm
$^{a)}$~Nikhef Theory Group, \\
Science Park 105, 1098 XG, Amsterdam, The Netherlands
\\
\vskip .1cm
$^{b)}$~Department of Physics and Astronomy, Vrije Universiteit Amsterdam,\\
1081 HV Amsterdam, The Netherlands
\\
\vskip .1cm
$^{c)}$Physik-Department, Technische Universit\"at M\"unchen\\
James-Franck-Str. 1, D-85748 Garching, Germany\\
TUM Institute of Advanced Study\\
Lichtenbergstr. 2a, D-85748 Garching, Germany
\vskip .1cm
$^{d)}$~Dipartimento di Fisica `G.~Galilei', Universit\`a di Padova\\
INFN, Sezione di Padova, Via Marzolo~8, I-35131 Padua, Italy\\
\end{center}

\vskip 0.3cm

\begin{abstract}
We consider an extension of the Standard Model in which the symmetry is enlarged by a global flavour factor $A_4$ and the scalar sector accounts for three copies of the Standard Model Higgs, transforming as a triplet of $A_4$. In this context, we study the most general scalar potential and its minima, performing for each of them a model independent analysis on the related phenomenology. We study the scalar spectrum, the new contributions to the oblique corrections, the decays of the $Z$ and $W^\pm$, the new sources of flavour violation, which all are affected by the introduction of multiple Higgses transforming under $A_4$. We find that this model independent approach discriminates the different minima allowed by the scalar potential.
\end{abstract}

\end{titlepage}

\vskip2truecm
%
%
\newpage

\section{Introduction}

The current data on neutrino oscillations seem to point at one small and two large angles in the neutrino mixing matrix \cite{FLMPR:NuData1,STV:NuData,MS:NuData, FLMPR:NuData2,GMS:NuData,KamLAND:NuData}. The data are consistent with various mixing patterns, where in particular the agreement with the tri-bimaximal \cite{HPS:TBM,Xing:TBM} mixing pattern is striking \cite{AF:ReviewDiscreteSyms}.

The use of non-Abelian discrete flavour symmetries has been proposed in different models (for a review see \cite{AF:ReviewDiscreteSyms}) to generate both the mentioned lepton mixing patterns and the quark ones. In general, in those models, one introduces so called flavons,  scalar fields charged in the flavour space, usually very heavy. Once the flavons develop specific vacuum expectation values (vevs), this translates to structures in the masses and mixings of the fermions.
However, imposing the correct symmetry breaking patterns on the flavons is highly non-trivial. This holds in particular if two or more flavons are used, breaking in different directions in flavour space. So far, only a few techniques have been developed, all of which need a supersymmetric context or the existence of extra dimensions \cite{AF:ReviewDiscreteSyms}.

Alternatively, one can look at models that require only one flavour symmetry  breaking  direction. In this case the scalar potential that implements the breaking can be non supersymmetric and does not require extra dimensions. Of particular interest is the possibility that one set of fields simultaneously takes the role of the flavons and the Standard Model (SM) Higgs fields, identifying the breaking scales of the electroweak and the flavour symmetries.

In this paper, we will consider the discrete flavour symmetry $A_4$ and we will assume that there are three copies of the Standard Model Higgs field, that transform among each other as a triplet of $A_4$ \cite{MR:A4EWscale,LK:A4EWscale,MP:A4EWscale,Ma:A4EWscale,Ma:TBfromA4alone,HMPV:DiscreteDM,MMP:DarkMatterA4}. The presence of this extended Higgs sector has an deep impact on the high energy phenomenology: indeed new contributions to the oblique corrections as well as new sources of flavour violation usually appear in this context. We will analyse the constraints coming from these observables for all the vacuum configurations allowed by the scalar potential and will discuss on the viability of each of them.

The structure of the paper is as follows. In section \ref{sec:potential}, we will introduce the scalar potential invariant under $A_4$ and under the gauge group of the Standard Model. In section \ref{sec:Higgs mix} we will introduce the various physical Higgs fields that are present in the model.

In the subsequent two sections we will present the different minima allowed by the potential and discuss the corresponding Higgs spectrum. These minima correspond to both real (section \ref{sec:realvevs}) and complex (section \ref{sec:complexvev}) vacuum expectation values of the Higgses. 

Section \ref{sec:bounds} we will discuss bounds on the allowed parameters using respectively unitarity constraints, decays of the $Z$ and $W^\pm$ bosons and constraints by oblique corrections. We note that all these bounds are rather model independent, meaning that they depend on the flavour symmetry assignment of the relevant Higgs fields, but not on those of the fermions in the theory. Further bounds can be derived from fermion decays and meson oscillations, but these bounds are always model dependent. We will present some of these in an accompanying paper \cite{ABMP:Constraining2}.

Finally, in section \ref{sec:results} we present the results of our analysis and in section \ref{sec:concl} we conclude. In the appendix A we report useful formulae for the analysis of the $T, S$ and $U$ parameters.

%
%
\section{The $A_4$ Scalar Potential}
\label{sec:potential}

We consider the Standard Model gauge group $SU(3)_c\times SU(2)_L\times U(1)_Y$ with the addition of a global flavour symmetry $A_4$ \cite{GR_Hamermesh,GR_Cornwell}. We consider three copies $\Phi_a$, $a = 1,2,3$, of the conventional SM Higgs field (i.e. a singlet of $SU(3)_c$, doublet of $SU(2)_L$ and with hypercharge $Y=1/2$) such that the three Higgses are in a triplet of the flavour group $A_4$. Once the flavour structure of the quarks and leptons is specified, each $\Phi_a$ will couple to the three fermion families according to the group theory rules in a model dependent way. We will study these couplings in more detail in \cite{ABMP:Constraining2}.

Below, we will write down the most general scalar potential for the three Higgses that is invariant under the flavour and gauge symmetries of the model. After the fields occupy one of the minima of the potential, electroweak symmetry gets broken (while electromagnetism is conserved) and we can develop the fields around their vacuum expectation values as
\be
\label{Higgs decomp}
\Phi_a =
\frac{1}{\sqrt{2}}
\bmat
\Re \; \Phi^{1}_a + i\, \Im \; \Phi^{1}_a \\
\Re \; \Phi^{0}_a + i\, \Im \; \Phi^{0}_a
\emat
\rightarrow \frac{1}{\sqrt{2}}
\bmat
\Re \; \phi_a^1 + i \, \Im \; \phi_a^1 \\
v_a e^{i \omega_a} + \Re \ \phi_a^0 + i \; \Im \, \phi_a^0
\emat.
\ee
Here $v_a e^{i \omega_a}$ is the vacuum expectation value of the $a^\textrm{th}$ Higgs field. One or two of the  $v_a$ can be zero, implying that the corresponding Higgs field does not develop a vev. Furthermore, if all vevs are real (so if all $\omega_a$ are zero) CP is conserved, while if one or more $\omega_a$s are nonzero, CP can be violated. Note that in general, there is the freedom to put one of the phases to zero by a global rotation.

We will use the $A_4$ basis as developed by Ma and Rajasekaran (MR) \cite{MR:A4EWscale}. The analysis could also be done in a different $A_4$ basis, for instance the one of Altarelli and Feruglio \cite{AF:Modular}. The results would then look different, but would obviously be equivalent. In the MR basis, the most general potential $V[\Phi_a] $ can be written as \cite{MR:A4EWscale,GH:2HDMReview}.
\beq
\label{A4pot}
\begin{split}
V[\Phi_a]=&\mu^2 (\Phi_1^\dag \Phi_1+ \Phi_2^\dag \Phi_2+ \Phi_3^\dag \Phi_3)+ \lambda_1 (\Phi_1^\dag \Phi_1+ \Phi_2^\dag \Phi_2+ \Phi_3^\dag \Phi_3)^2+\\
&+ \lambda_3 (\Phi_1^\dag\Phi_1 \Phi_2^\dag\Phi_2+ \Phi_1^\dag\Phi_1\Phi_3^\dag \Phi_3+ \Phi_2^\dag\Phi_2\Phi_3^\dag \Phi_3)+ \\
&+ \lambda_4 (\Phi_1^\dag \Phi_2 \Phi_2^\dag \Phi_1 +\Phi_1^\dag \Phi_3 \Phi_3^\dag \Phi_1+ \Phi_2^\dag \Phi_3 \Phi_3^\dag \Phi_2)+\\
&+\frac{\lambda_5}{2}\bigg[ e^{i \epsilon} \left[ (\Phi_1^\dag \Phi_2)^2+(\Phi_2^\dag \Phi_3)^2+(\Phi_3^\dag \Phi_1)^2\right]+  e^{-i \epsilon}   \left[(\Phi_2^\dag \Phi_1)^2+(\Phi_3^\dag \Phi_2)^2+(\Phi_1^\dag \Phi_3)^2\right] \bigg] \,,
\end{split}
\eeq
in agreement with the usual notation adopted in the two Higgs Doublet Models (2HDM).
The parameter $\mu^2$ is typically negative in order to have a stable minimum away from the origin. All the other parameters, $\lambda_i$, are real parameters which must undergo to the condition for a potential bounded from below: this forces $\lambda_1$ and the combination $\lambda_1 + \lambda_3 + \lambda_4 + \lambda_5 \cos \epsilon$ to be positive.

It is interesting to notice that, contrary to other multi Higgs (MH) scenarios, here we can not recover the SM limit, with one light scalar and all the others decoupled and very heavy. The flavour symmetry constrains the potential parameters in such a way that  the scalar masses are never independent from each other. This can be easily understood by a parameter counting: the scalar potential in eq. (\ref{A4pot}) presents $6$ independent parameters and the number of the physical quantities is $8$, i.e. the electroweak (EW) vev and the seven masses for the massive scalar fields.

We will study the minima of the potential in eq. (\ref{A4pot}) under electromagnetism conserving vevs as specified in eq. (\ref{Higgs decomp}) by studying the first derivative system
\be
\label{first deriv gen}
\frac{\partial V[\Phi]}{\partial  \Phi_{\mathcal{I}}}=0\,,
\ee
where $\Phi_{\mathcal{I}}$ is of the fields $\Re \; \Phi_a^1$, $\Re \; \Phi_a^0$, $\Im \; \Phi_a^1$ or $\Im \; \Phi_a^0$ and by requiring that the Hessian
\be
\frac{\partial^2 V[\Phi]}{\partial  \Phi_{\mathcal{I}} \partial  \Phi_{\mathcal{J}}}
\ee
has non negative eigenvalues, or in other words that all the physical masses are positive except those ones corresponding to the Goldstone bosons (GBs) that vanish.

In sections \ref{sec:realvevs} and \ref{sec:complexvev} we will verify that this potential presents a number of solutions. Some of them are natural in the sense that they do not require \emph{ad hoc} values of the potential parameters; these are only constrained by requiring the boundness at infinity and the positivity of all the physical scalar masses. The only potential parameter constrained  is the bare mass term $\mu^2$ which is related to the physical  Electroweak (EW) vev, $v_w^2 = v_1^2+v_2^2+v_3^2$. Others require specific relations between the adimensional scalar potential parameters and may have extra Goldstone bosons.

%
%
\section{The Physical Higgs Fields}
\label{sec:Higgs mix}
The symmetry breaking of the Higgs fields of equation eq. (\ref{Higgs decomp}) leads to a large number of charged and neutral Higgs bosons as well as the known Goldstone bosons of the Standard Model.

In the most general case, where CP is not conserved, the neutral real and imaginary components of eq. (\ref{Higgs decomp}) mix to five CP non-definite states and a GB:
\beq
\label{defmat}
\ba{rcl}
h_\alpha &=& U_{\alpha a} \Re \, \phi_a^0 + U_{\alpha (a+3)} \Im \, \phi_a^0\,,\\[2mm]
\pi^0&=&U_{6 a} \Re \, \phi_a^0 + U_{6 (a+3)} \Im \, \phi_a^0\,.
\ea
\ee
Here $a = 1,2,3$ and $\alpha = 1 - 5$, while $\alpha = 6$ represents the GB $\pi^0$. In matrixform this reads
\beq
\label{defU}
\bmat h_1 \\ \vdots \\ h_5 \\ \pi^0 \emat = U \bmat \Re \, \phi_1^0 \\ \vdots \\ \Re \, \phi_3^0 \\ \Im \, \phi_1^0 \\ \vdots \\ \Im \, \phi_3^0 \emat
\eeq
Clearly \eq{defmat} holds also in the CP conserved case: in that case the 6 by 6  scalar mass matrix reduces to a block diagonal matrix with two 3 by 3 mass matrices leading to three CP even states and 2 CP odd states and the GB $\pi^0$.

The three charged scalars mix into two new charged massive states and a charged GB.
\beq
\label{defS}
\bmat H^+_{1} \\ H^+_{2} \\ \pi^+  \emat = S\, \bmat \phi^1_1 \\ \phi^1_2 \\ \phi^1_3 \emat\,,
\eeq
where $\pi^+$ is the Goldstone boson eaten by the gauge bosons $W^+$. In general, the  $S$ is a complex unitary matrix. In the special case where CP is conserved, its entries are real (and it is thus an orthogonal matrix).

%
%
\section{Solutions with real vevs}
\label{sec:realvevs}

In this section, we will study minima of the potential in eq.~(\ref{A4pot}) in which only $\Re\,\phi^0_a$ develops a vev, i.e. the vev is real and the CP symmetry is conserved. In this case we expect having 3 neutral scalar CP-even states, 2 CP-odd states and 2 charged scalars as well as a real and a complex GBs originating from respectively the CP-odd states and the charged states.

In this case, all the $\omega_a$ vanish and the first derivative system in eq. (\ref{first deriv gen}) reduces to
\beq
\label{firstder}
\ba{rcl}
v_1 [ 2 (v_1^2+v_2^2 +v_3^2)\lambda_1+ (v_2^2+v_3^2) (\lambda_3+\lambda_4+\lambda_5 \cos\epsilon)+2 \mu^2] &=&0\,,\\[2mm]
v_2 [ 2 (v_1^2+v_2^2 +v_3^2)\lambda_1+ (v_1^2+v_3^2) (\lambda_3+\lambda_4+\lambda_5 \cos\epsilon)+2 \mu^2] &=&0\,,\\[2mm]
v_3 [ 2 (v_1^2+v_2^2 +v_3^2)\lambda_1+ (v_1^2+v_2^2) (\lambda_3+\lambda_4+\lambda_5 \cos\epsilon)+2 \mu^2] &=&0\,,\\[2mm]
v_1(v_2^2-v_3^2)\lambda_5 \sin\epsilon &=&0\,,\\[2mm]
v_2(v_1^2-v_3^2)\lambda_5 \sin\epsilon &=&0\,,\\[2mm]
v_3(v_2^2-v_1^2)\lambda_5 \sin\epsilon &=&0\,,
\ea
\ee
where the first three derivatives refer to the real components $\Phi_a^0$ and the second ones to the imaginary parts. In the most general case, when neither $\epsilon$ nor $\lambda_5$ is zero, the last three  equations allow two different solutions
\begin{itemize}
\item[1)] $v_1=v_2=v_3=v=v_w/\sqrt{3}$;
\item[2)] $v_1 \neq 0$ and $v_2=v_3=0$ (and permutations of the indices); in this case $v_1=v_w$.
\end{itemize}
Both these solutions are solutions of the first three equations as well, provided that
\beq
\begin{cases}
\mu^2=-( 3 \lambda_1+\lambda_3 +\lambda_4+\lambda_5\cos\epsilon)v_w^2/3 & \text{for the first case}\\
\mu^2=-\lambda_1 v_w^2 & \text{for the second case.}
\end{cases}
\eeq
In this cases $\lambda_5$ can be chosen positive, as a sign can be absorbed in a redefinition of $\epsilon$.

Next, we consider the case where $\sin \epsilon$ is 0. This implies $\epsilon = 0$ or $\pi$. We may however absorb the minus sign corresponding to the second case in a redefinition of $\lambda_5$ that is now allowed to span over both positive and negative values.

Assuming $v_1 \neq 0$, we may solve the first equation in \eq{firstder} with respect to $\mu^2$. Then by substituting  $\mu^2$  in the other two equations we get
\beq
\ba{rcl}
v_2(v_1^2-v_2^2)(\lambda_3+\lambda_4 + \lambda_5)&=&0\,,\\[2mm]
v_3(v_1^2-v_3^2)(\lambda_3+\lambda_4 + \lambda_5)&=&0\,.
\ea
\ee
Next to the two solutions present in the general case, this system has two further possible solutions
\begin{itemize}
\item[3)]$v_3=0,v_2=v_1=v_w/\sqrt{2}$ and permutations. This requires
\beq
\mu^2 = -\left(4\lambda_1 + \lambda_3 + \lambda_4 + \lambda_5\right)v^2_w/4\,.
\eeq
\item[4)]$(\lambda_3+\lambda_4 + \lambda_5)$ = 0. This condition implies that in the real neutral direction there is a enlarged--$O(3)$ accidental symmetry that is spontaneously broken by the vacuum configuration, thus we xpect extra GBs.  Indeed  in this case $v_1$, $v_2$ and $v_3$ are only restricted to satisfy  $v_1^2+v_2^2+v_3^2=v_w^2$ and the parameter $\mu^2$ is given by $\mu^2 = - \lambda_1 v^2_w $.
\end{itemize}

Finally, the case $\lambda_5=0$ allows special cases of the solutions 1) to 4), but does not give rise to new solutions. For this reason, we will discuss only the general cases and the case $\epsilon=0$ in the remainder of this section and comment what happens for $\lambda_5=0$.

\mathversion{bold}
\subsection{$\epsilon\neq0$: The Alignment $(v,v,v)$}
\label{secvvv}
\mathversion{normal}

In the basis chosen, the vacuum alignment $(v,v,v)$ preserves the $Z_3$ subgroup of $A_4$\footnote{In the special case where $\epsilon = 0$, the symmetry of the vacuum is enlarged to $S_3$ even if $S_3$ is not a subgroup of $A_4$. The reason is that setting $\epsilon = 0$ effectively enlarges the symmetry of the potential to $S_4$ (once also $SU(2) \times U(1)$ gauge invariance is required), which does have $S_3$ as a subgroup.}.  It is convenient to perform a basis transformation into the $Z_3$ eigenstate basis,  $1,1'\sim \omega,1''\sim \omega^2$   according to
\bea
\label{Z3eig}
\varphi&=& (\Phi_1+\Phi_2+\Phi_3)/\sqrt{3} \sim 1 \nn\\
\varphi'&=& (\Phi_1+\omega \Phi_2+\omega^2 \Phi_3)/\sqrt{3} \sim \omega \nn\\
\varphi''&=& (\Phi_1+\omega^2 \Phi_2+\omega \Phi_3)/\sqrt{3}  \sim\omega^2\,.
\eea
When $A_4$ is broken to $Z_3$ in the $Z_3$ eigenstate basis, $\varphi\sim 1$ behaves like the standard Higgs doublets: its  neutral real component  develops a vacuum expectation values $\vev{\varphi^{0R}}=v_w$  and all its other components correspond to the GBs eaten by the corresponding gauge bosons. The physical real scalar gets a mass given by
\be
m_{h_1}^2=\frac{2}{3} v_w^2 ( 3 \lambda_1+  \lambda_3+ \lambda_4+\lambda_5\cos\epsilon).
\ee
The neutral components of the other two doublets $\varphi'$ and $\varphi''$ mix into two complex neutral states and their masses are given by
\beq
\label{massp}
m^{\prime,\prime\prime\,2}_{n}=\dfrac{v_w^2}{6}\left(-\lambda_3-\lambda_4-4 \lambda_5 \cos \epsilon \pm \sqrt{(\lambda_3+\lambda_4)^2+4\lambda_5^2(1+2\sin^2\epsilon)-4(\lambda_3+\lambda_4)\lambda_5\cos\epsilon}\right)\,.
\ee
The charged components of $\varphi',\varphi''$ do not mix, their masses being
\bea
\label{massch}
m^{\prime,\prime\prime\,2}_{ch}=-\frac{v_w^2}{6}  \left(3 \lambda_4+3
   \lambda_5 \cos \epsilon \pm\sqrt{3}
   \lambda_5 \sin \epsilon \right)\,.
\eea

\mathversion{bold}
\subsection{$\epsilon\neq0$: The Alignment $(v,0,0)$}
\label{secv00}
\mathversion{normal}

In the chosen $A_4$ basis, the vacuum alignments $(v,0,0)$ preserves the $Z_2$ subgroup of $A_4$.  As we did with the vacuum alignment that conserved the $Z_3$ subgroup,  in this case it is useful to rewrite the scalar potential by performing the following $Z_2$ conserving basis transformation
\beq
\ba{rcl}
\Phi_1&\to& \Phi_1\,,\\[2mm]
\Phi_2&\to & e^{-i \epsilon/2} \Phi_2\,,\\[2mm]
\Phi_3&\to & e^{i \epsilon/2} \Phi_3\,.
\ea
\ee
$\Phi_1$  is even under $Z_2$ and behaves like the standard Higgs doublet, while $\Phi_2$ and $\Phi_3$ are odd.
For what concerns the neutral states, the $6\times6$ mass matrix is diagonal in this basis and with some degenerated entries: using a notation similar to the 2DHM, we have
\beq
\ba{ll}
m^2_{h_1}\equiv2\lambda_1v_w^2\,,\qquad\qquad
&m^2_{h_2}=m^2_{h_3} =\dfrac{1}{2} (\lambda_3+\lambda_4-\lambda_5 ) v_w^2\,,\\[2mm]
m^2_{h_4}=m^2_{h_5} =\dfrac{1}{2} (\lambda_3+\lambda_4+\lambda_5 ) v_w^2\,,\qquad\qquad
&m^2_{\pi^0}=0\;,
\ea
\ee
where the last state corresponds to the GB.
The charged scalar mass matrix is also diagonal with
\beq
m^2_{C_1}=m^2_{C_2}=\dfrac{1}{2} \lambda_3 v_w^2\,,\qquad\qquad
m^2_{\pi^+}=0\,,
\ee
where the last state corresponds to the GB. The degeneracy in the mass matrices are imposed by the residual $Z_2$ symmetry. Contrary to the previous case the neutral scalar mass eigenstates are real and not complex.

\mathversion{bold}
\subsection{$\epsilon=0$: The Alignment $(v,v,0)$}
\mathversion{normal}
\label{realvevssub}

This vacuum alignment does not preserve any subgroup of $A_4$ and it holds that $v=v_w/\sqrt{2}$. From the minimum equations we have that
\be
\mu^2=-\frac{1}{4}v_w^2 (4 \lambda_1+\lambda_3+\lambda_4+\lambda_5)\,.
\ee
The scalar and pseudoscalar mass eigenvalues are given by
\beq
\label{MIvv0}
\ba{ll}
m_{h_1}^2=-\dfrac{v_w^2}{2} ( \lambda_3+\lambda_4+\lambda_5)\,,\qquad\qquad
&m_{h_2}^2=\dfrac{v_w^2}{2} (4\lambda_1+ \lambda_3+\lambda_4+\lambda_5)\,,\\[2mm]
m_{h_3}^2=\dfrac{v_w^2}{4} ( \lambda_3+\lambda_4+\lambda_5)\,,\qquad\qquad
&m_{h_4}^2=-\lambda_5 v_w^2 \,,\\[2mm]
m_{h_5}^2=\dfrac{v_w^2}{4} (\lambda_3+\lambda_4-3\lambda_5 )\,,\qquad\qquad
&m_{\pi^0}^2=0\,.
\ea
\ee
For the charged sector we have
\be
\label{MCvv0}
m^2_{C_1}=\dfrac{v_w^2}{4}(\lambda_3-\lambda_4-\lambda_5 )\,,\qquad\qquad
m^2_{C_2}=-\dfrac{v_w^2}{2}(\lambda_4+\lambda_5 )\,\qquad\qquad
m^2_{C_3}=0\,.
\ee

For $\lambda_5\neq0$ the alignment $(v,v,0)$ has the correct number of GBs, while for $\lambda_5=0$ we have an extra massless pesudoscalar. However in both cases, $\lambda_5\neq0$ or $\lambda_5=0$, the conditions $m_{h_1}^2>0$ and $m_{h_3}^2>0$ can not be simultaneously satisfied. This alignment is therefore a saddle point of the $A_4$ scalar potential we are studying.

\mathversion{bold}
\subsection{$\epsilon=0$: The Alignment $(v_1,v_2,v_3)$}
\mathversion{normal}
\label{secv1v2v3}

This vacuum alignment, as the previous one, does not preserve any subgroup of $A_4$. A part from the condition $\epsilon=0$, we recall that in this case there is the further constraint $\lambda_3+\lambda_4+\lambda_5=0$ and $\lambda_5$ may assume both positive and negative values since we have reabsorbed in the $\lambda_5$ sign the case $\epsilon=\pi$.

Let us define $v_w^2= v_1^2+v_2^2+v_3^2= (1+s^2+r^2)v_1^2$ with  $s= v_2/v_1$ and $r= v_3/v_1$  respectively.
The mass matrix for the neutral scalar states presents two null eigenvalues--as we expected since the condition $\lambda_3+\lambda_4+\lambda_5=0$  enlarges the potential symmetry--  and a massive one
\beq
m{h_1}^2= 2 \lambda_1 v_w^2\,.
\eeq
At the same time the mass matrix for the CP-odd states has one null eigenvalue--the GB $\pi^0$ and two degenerate eigenvalues of mass
\beq
m^2_{h_2}=m^2_{h_3}=(\lambda_3+\lambda_4)v_w^2\,.
\eeq
Notice that for the special case $\lambda_5=0$ we have the constraint $\lambda_3=-\lambda_4$ that implies two extra massless pseudoscalars. Finally for the charged scalars we have
\beq
\ba{l}
m^2_{C_1}=m^2_{C_2}=\dfrac{1}{2}\lambda_3 v_w^2\,,\qquad \qquad
m^2_{C3}=0
\ea
\ee
The total amount of GBs is 5 (7) for the case $\lambda_5\neq0$ ($\lambda_5=0$), so we have 2 (4) extra unwanted GBs: this situation is really problematic. We note that the introduction of terms in the potential that softly break $A_4$ can ameliorate the situation with the Goldstone bosons. We will analyse soft $A_4$ breaking terms in more detail in \cite{ABMP:Constraining2}.

%
%
\section{Solutions with complex vevs}
\label{sec:complexvev}

In this subsection, we consider vacua that exhibit complex vevs. In general this could lead to spontaneous CP violation and we will comment in section \ref{secCPV} whether this is the case for the solutions we discuss. We recall that a global rotation can always absorb one of the three phases of the vevs.

We note that the two natural vacua of the previous section $(v,v,v)$ and $(v,0,0)$ do not have complex analogues, as they have only one phase that can be reabsorbed.

\mathversion{bold}
\subsection{ The Alignment $(v_1 e^{i \omega_1}, v_2, 0)$}
\label{subsecvomegav0}
\mathversion{normal}

In this case the third doublet is inert and therefore we are left only with two doublets that develop a complex vev and after the redefinition, there is only one phase $\omega_1$. Taking the generic solution $(v_1 e^{i \omega_1},v_2  ,0)$ the minimum equations are given by

\beq
\label{firstderno}
\ba{l}
v_1 \left [ \cos\omega_1 [2 \mu^2 + 2\lambda_1 (v_1^2+v_2^2) +(\lambda_3+\lambda_4)v_2^2]  +\lambda_5  v_2^2\cos (\epsilon+\omega_1)]\right ]
=0\,,\\[2mm]
v_2 \left [   (2 \mu^2 + 2\lambda_1 (v_1^2+v_2^2) +(\lambda_3+\lambda_4)v_1^2 +\lambda_5  v_1^2\cos (\epsilon +2  \omega_1)\right ]=0\,,\\[2mm]
\\
v_1 \left [ \sin\omega_1 [2 \mu^2 + 2\lambda_1 (v_1^2+v_2^2) +(\lambda_3+\lambda_4)v_2^2]-\lambda_5  v_2^2\sin(\epsilon+\omega_1) \right]
=0\,,\\[2mm]
v_2 v_1^2\sin (\epsilon+2  \omega_1)=0\,.
\ea
\ee
The last equation can be solved by $\epsilon=-2 \omega_1$ or $\epsilon=-2 \omega_1 + \pi$. Like in section \ref{sec:realvevs}, we can absorb the second case by a redefinition of $\lambda_5$. The other three equations reduce to
\beq
\label{firstdernob}
\ba{l}
v_1 \cos\omega_1 \left [  2 \mu^2 + 2\lambda_1 (v_1^2+v_2^2) +(\lambda_3+\lambda_4)v_2^2 + \lambda_5  v_2^2 \right ]=0\,,\\[2mm]
v_2 \left [   2 \mu^2 + 2\lambda_1 (v_1^2+v_2^2) +(\lambda_3+\lambda_4)v_1^2 + \lambda_5  v_1^2\right ]=0\,,\\[2mm]
v_1 \sin\omega_1  \left [  2 \mu^2 + 2\lambda_1 (v_1^2+v_2^2) +(\lambda_3+\lambda_4)v_2^2 + \lambda_5  v_2^2  \right]=0\,.
\ea
\ee
that are simultaneously solved for $v_1=v_2=v_w/\sqrt{2}$ and
\be
\mu^2=-\frac{v_w^2}{4}( 4\lambda_1+\lambda_3+\lambda_4 + \lambda_5)\,.
\ee
The neutral and charged $6\times 6$ mass matrices are quite simple and it is possible having analytical expression for the mass eigenvalues. For the neutral sector we have
\be
\begin{array}{ll}
 m_{h_1}^2=\dfrac{1}{2} v_w^2(-\lambda_3-\lambda_4 - \lambda_5)\,,\qquad\qquad
 &m_{h_2}^2= \dfrac{1}{2} v_w^2 (4 \lambda_1+\lambda_3+\lambda_4 + \lambda_5)\,,\\[2mm]
 m_{h_3}^2=\dfrac{1}{4} v_w^2 (\lambda_3+\lambda_4 - \lambda_5 + 2 \lambda_5 \cos 3 \omega_1)\,,\qquad\qquad
 &m_{h_4}^2= - \lambda_5 v_w^2\,,\\[2mm]
 m_{h_5}^2= \dfrac{1}{4} v_w^2 (\lambda_3+\lambda_4 - \lambda_5 - 2 \lambda_5 \cos 3 \omega_1)\,,\qquad\qquad
 &m_{\pi^0}^2=0\,,
 \end{array}
\ee
and for the charged one we have
\be
m^2_{C_1}=\dfrac{v_w^2}{4}(\lambda_3-\lambda_4 - \lambda_5 )\,,\qquad\qquad
m^2_{C_2}=\dfrac{v_w^2}{2}(-\lambda_4 - \lambda_5 )\,,\qquad\qquad
m^2_{C_3}=0\,.
\ee
We see that the mass of the fourth neutral boson selects negative values for $\lambda_5$, i.e.  the second solution  $\epsilon=-2 \omega_1 + \pi$.
It is interesting to see that in the limit $\omega_1 \rightarrow 0 $ (or $\pi$), it is not possible to have both $m^2_{h_1}$ and $m^2_{h_3}$ (respectively $m^2_{h_5}$) positive, but that in the general case, there are points in parameter space where indeed all masses are positive. This is in particular clear in the region around $\cos 3 \omega_1 = 0$.

Finally, as for the case with only real vevs, for $\lambda_5=0$ we have two problems: an extra GB and  we cannot have all positive massive eigenstates.

\mathversion{bold}
\subsection{The Alignment $(v_1 e^{i\omega_1}, v_2 e^{i\omega_2}, v_3)$}
\mathversion{normal}
\label{seccomplexvev2}

In this case all the doublets  develop a vev $v_i\neq 0$, so we may have two physical phases. We have the freedom to take $\omega_3=0$. In this case the first derivatives system is given by

\beq
\ba{l}
\begin{split}
v_1 &\left \{ \cos\omega_1 [2 \mu^2 + 2\lambda_1 (v_1^2+v_2^2+v_3^2) +(\lambda_3+\lambda_4)(v_2^2+v_3^2)]+\right.\\
 &\qquad \qquad\left.+\lambda_5 [v_3^2\cos (\epsilon-\omega_1)+ v_2^2\cos (\epsilon+\omega_1-2 \omega_2)] \right \} = 0\,,
\end{split}\\[2mm]
\begin{split}
v_2 &\left \{ \cos\omega_2 (2 \mu^2 + 2\lambda_1 (v_1^2+v_2^2+v_3^2) +(\lambda_3+\lambda_4)(v_1^2+v_3^2)+\right.\\
& \qquad \qquad\left.+\lambda_5 [v_3^2\cos (\epsilon+\omega_2)+ v_1^2\cos (\epsilon-\omega_2+2  \omega_1)]\right \}=0\,,
\end{split}\\[2mm]
\begin{split}
v_3 &\left \{ 2 \mu^2 + 2\lambda_1 (v_1^2+v_2^2+v_3^2) +(\lambda_3+\lambda_4)(v_1^2+v_2^2)+\right.\\
& \qquad \qquad\left.+\lambda_5 [v_1^2\cos (\epsilon-2\omega_1)+ v_2^2\cos (\epsilon+2  \omega_2)]\right \}=0\,,
\end{split}\\[2mm]
\begin{split}
v_1 &\left \{ \sin\omega_1 [2 \mu^2 + 2\lambda_1 (v_1^2+v_2^2+v_3^2) +(\lambda_3+\lambda_4)(v_2^2+v_3^2)]+\right.\\
& \qquad \qquad\left.+\lambda_5 [v_3^2\sin (\epsilon-\omega_1)- v_2^2\sin(\epsilon+\omega_1-2 \omega_2)]\right\}=0\,,
\end{split}\\[2mm]
\begin{split}
v_2 &\left \{ \sin\omega_2 (2 \mu^2 + 2\lambda_1 (v_1^2+v_2^2+v_3^2) +(\lambda_3+\lambda_4)(v_1^2+v_3^2))+\right.\\
& \qquad \qquad\left.+\lambda_5 [-v_3^2\sin (\epsilon+\omega_2)+ v_1^2\sin (\epsilon-\omega_2+2  \omega_1)]\right\}=0\,,
\end{split}\\[2mm]
v_3 \left[\lambda_5 (- v_1^2\sin (\epsilon-2\omega_1)+ v_2^2\sin (\epsilon+2  \omega_2))\right]=0\,.
\ea
\ee
The last equation is solved for $\omega_2=-\omega_1$ and $ v_2=v_1=v$. Defining $v_3=r v$ and $v_1^2 + v_2^2 + v_3^2=v_w^2$   the previous system reduces to the three equations
\beq
\label{dirsderno2}
\ba{l}
\mu^2+\dfrac{v_w^2}{2(2+r^2)}\left[ (4+ 2 r^2) \lambda_1+(1+r^2) (\lambda_3+\lambda_4)+\dfrac{\lambda_5}{\cos\omega_1} ( r^2 \cos(\epsilon-\omega_1)+  \cos(\epsilon+3\omega_1))\right]=0\,,\\[2mm]
\mu^2+\dfrac{v_w^2}{2(2+r^2)}\left[ (4+ 2 r^2) \lambda_1+(1+r^2) (\lambda_3+\lambda_4)+\dfrac{\lambda_5}{\sin\omega_1} ( r^2 \sin(\epsilon-\omega_1)+  \sin(\epsilon+3\omega_1))\right]=0\,,\\[2mm]
\mu^2+\dfrac{v_w^2}{(2+r^2)}\Big[ (2+  r^2) \lambda_1+ \lambda_3+\lambda_4+\lambda_5 \cos(\epsilon-2 \omega_1)\Big]=0\,.
\ea
\ee

We can solve the third equation in  \eq{dirsderno2}  in terms of $\mu^2$ and then the second   equation in terms of $\lambda_5$, giving
\beq
\label{solgen}
\ba{rcl}
\mu^2&=&-\frac{v_w^2}{2+r^2} [(2+r^2)\lambda_1+\lambda_3+\lambda_4+\lambda_5 \cos(\epsilon-2 \omega_1)] \,,\\[2mm]
\lambda_5 &=&  \dfrac{(r^2-1)(\lambda_3+\lambda_4)\sin{\omega_1}}{(r^2-1)\sin(\epsilon-\omega_1)-2\cos\epsilon\sin(3\omega_1)}\,.
\ea
\ee
Then the first equation in \eq{dirsderno2} has two possible solutions, for $\lambda_4$ and $\epsilon$ respectively
\beq
\label{teps}
\ba{ll}
i)& \lambda_4=-\lambda_3 \,,\\[2mm]
ii) & \tan\epsilon=\dfrac{r^2 \sin 2 \omega_1+ \sin 4 \omega_1}{r^2 \cos 2 \omega_1- \cos 4 \omega_1}\,.
\ea
\ee

To test the validity of the solution so far sketched it is necessary to check to be in a true minimum of the potential and   not to have  extra GBs a part from three corresponding to the GBs eaten by the gauge bosons. However the  relations given in \eq{solgen} and \eq{teps} do not allow to get analytical solutions for the scalar masses in case $ii)$. For this reason we will consider only three special limits in this case : $r\sim0$, $r\sim1$ and $r$ very large. We think that these limit situations could be the most interesting ones in  the model building realizations. Indeed  models present in literature \cite{LK:A4EWscale,MP:A4EWscale} fall in the third case, $r$ very large.

\subsubsection{Case $i)$ }
\label{seccomplecvevscasei}

In this case the constraints $ \lambda_4=-\lambda_3$ puts $\lambda_5$ to zero and enlarge substantially the symmetries of the potential: we have an accidental $O(3)$ in the neutral real direction  and two accidental $U(1)$s due to $\lambda_5=0$. For this reason the neutral spectrum has 5 massless particles, the GB $\pi^0$ and 4 other GBs, and only one massive state
\beq
m_{h_1}^2= 2\lambda_1 v_w^2\,.
\eeq
The charged scalars are
\beq
\ba{l}
m^2_{C_1}=m^2_{C_2}=\dfrac{1}{2}\lambda_3 v_w^2\,,\qquad \qquad
m^2_{C3}=0
\ea
\ee
The massive states are degenerate as in the case with real vevs studied in \sec{realvevssub} for $\lambda_5=0$.

\subsubsection{Case $ii)$ }
\label{seccomplecvevscaseii}

As it is not possible to find analytical solutions, here we will study three special limits of case ii.

\noindent
\textbf{$\bullet$\quad$r \sim0$}

\noindent In this case we will neglect terms of order $r^2$. From \eq{teps}  we have that for $r\sim 0$
\be
\epsilon\sim - 4 \omega_1+ N \pi\,,
\ee
thus from \eq{solgen} we have
\bea
\mu^2 &=&-\lambda_1 v_w^2 - (\lambda_3+\lambda_4) \frac{1-\cos 6\omega_1}{2- 4 \cos 6 \omega_1}\,,\nn\\
\lambda_5 &=& \frac{\lambda_3+\lambda_4}{1- 2 \cos 6 \omega_1}\,.
\eea

Under these approximations the 6 x  6 neutral scalar mass matrix gives one null mass state, $m^2_{\pi^0}=0$,  corresponding to the GB and  the following five eigenvalues at leading order, given  by
\be
\label{eqneutral}
\begin{array}{l}
m^2_{h_1}\sim f[\lambda_i] \mathcal{O} (r^2) v_w^2\\[2mm]
 m^2_{h_2}\sim  -(\lambda_3+\lambda_4)/(1-2 \cos 6 \omega_1) v_w^2\,\\[2mm]
m^2_{h_3}\sim  \left[-2\lambda_1 + (4\lambda_1+\lambda_3+\lambda_4)(1-\cos 6\omega_1)/(1-2 \cos 6 \omega_1)\right]v_w^2 \\[2mm]
m^2_{h_4}\sim -\left[(\lambda_3+\lambda_4)\cos 6\omega_1 v_w^2/(1-2 \cos 6 \omega_1)\right] v_w^2\,,\\[2mm]
m^2_{h_5}\sim   -\left[2(\lambda_3+\lambda_4)\sin^2 3\omega_1 /(1-2 \cos 6 \omega_1)\right] v_w^2\,,
\end{array}
\ee
where  $f[\lambda_i]$ stays for a linear combination of the adimensional $\lambda$ parameters of the potential. The previous neutral spectrum present a lightest state that may be too light to be phenomenologically acceptable. Assuming that the $\lambda$'s potential parameters run in the `natural' range $0.1\div10$ or, somewhat optimistically, $10^{-2}\div10^2$.  For what concerns $r$ we are in the limit of $r^2\sim 0$, so as reference value we may take $r^2\sim 10^{-3}\div10^{-2}$. By combining these two ranges we find upper bounds
\bea
m^2_{h_1}\leq 200 \, \mbox{GeV} &\quad& \mbox{for}\quad \lambda_i\sim 100, r^2\sim 10^{-2}\,,\nn\\
m^2_{h_1}\leq 25 \, \mbox{GeV} &\quad& \mbox{for}\quad \lambda_i\sim 10, r^2\sim 10^{-3}\,.
\eea
Since  $f[\lambda_i]\sim 100$ may be obtained only for very peculiar combinations of the potential parameters, the previous estimates indicate that for relative tiny value of $r$ the spectrum may present very  light neutral states.

On the contrary, in the charged sector we have the two  GBs eaten by the corresponding gauge bosons, $m_{C_3}^2=0$,  and  two complex massive states with masses
\be
\begin{array}{l}
m_{C_1}^2\sim -[ \lambda_4+(\lambda_3+ \lambda_4 \cos6 \omega_1)/(1-2 \cos6 \omega_1)]v_w^2/2  \\[2mm]
 m_{C_2}^2\sim -[2 \lambda_4+(\lambda_3+2 \lambda_4 \cos6 \omega_1)/(1-2 \cos6 \omega_1)]v_w^2/2\,.
\end{array}
\ee

\noindent
\textbf{$\bullet$\quad$r \sim1$}

In this limit we may write $r\sim 1+\delta$ and make an expansion in terms of $\delta$ neglecting terms of order $\delta^2$. Thus we have
\be
\epsilon\sim \pi/2- \omega_1-\delta \cot 3 \omega_1+ N \pi \,,
\ee
and then
\bea
\mu^2 &=&-(3\lambda_1+\lambda_3+\lambda_4)/3 v_w^2 - \delta/9 (\lambda_3+\lambda_4) v_w^2\,,\nn\\
\lambda_5 &=& \delta (\lambda_3+\lambda_4)\csc 3 \omega_1\,.
\eea

Under these approximations the 6 x  6 neutral scalar mass matrix gives the usual  null mass state, $m^2_{\pi^0}$, corresponding to the GB and  the following five eigenvalues
\be
\label{eqneutral}
\begin{array}{l}
m^2_{h_1}\sim m^2_{h_2} \sim f[\lambda_i] \mathcal{O} (\delta^2)  v_w^2\,,\\[2mm]
m^2_{h_3}\sim m^2_{h_4} \sim - (\lambda_3+\lambda_4)/3 v_w^2 \,,\\[2mm]
m^2_{h_5}\sim  2(3 \lambda_1+\lambda_3+\lambda_4)/3 v_w^2\,,
\end{array}
\ee
where again   $f[\lambda_i]$ stays for a linear combination of the $\lambda$'s potential parameters. A analysis similar to the one for the case with $r\sim0$ shows that the neutral spectrum may present very light states.

In the charged sector we have the  GBs  eaten by the gauge bosons and   two degenerate massive state
\be
m_{C_1}^2\sim m_{C_2}^2\sim -\lambda_4/2 v_w^2\,.
\ee

\noindent
\textbf{$\bullet$\quad$r \gg1$}

In this  case we may perform an expansion in term of $1/r$ and neglect terms of order $1/r^2$.  From \eq{teps} we have that
\be
\epsilon\sim 2 \omega_1+N \pi\,,
\ee
and then \eq{solgen} reduces to
\bea
\mu^2&\sim& - \lambda_1 v_w^2  \,,\nn\\
\lambda_5&\sim& -(\lambda_3+\lambda_4) \,,\nn\\
\eea
Under these approximations we find a massless neutral scalar state, $m^2_{\pi^0}=0$, and the other 5 neutral masses are given at leading order by
\beq
\label{eqneutral}
\ba{l}
m^2_{h_1}\sim m^2_{h_2} \sim f[\lambda_i] \mathcal{O} (1/r^2)  v_w^2\,,\\[2mm]
m^2_{h_3}\sim   2 \lambda_1 v_w^2\,,\\[2mm]
m^2_{h_4}\sim m^2_{h_5}\sim   (\lambda_3+\lambda_4) v_w^2\,,
\ea
\ee
where once more   $f[\lambda_i]$ stays for a linear combination of the $\lambda$'s potential parameters. The charged scalar mass matrix is diagonal up to terms of order $\mathcal{O}(1/r^2)$ with  two massive  degenerate states
\be
m^2_{C_1}=m^2_{C_2}=\lambda_3 {v_w^2}/{2 }\,,
\ee
and the correct number of GBs.

If we consider now \eq{eqneutral} we see that as for $r\sim 0$  and $r\sim1$ the expressions for $m^2_{h_{1,2}}$ say that we may have  two very light neutral scalars.  Taking  as reference values for $r$ the range $50\div200$ we find
\be
m^2_{h_{1,2}}\sim \sqrt{f[\lambda_{i}}]  \,5 \, \mbox{GeV} ( 1\,  \mbox{GeV})\,,
\ee
giving
\bea
m^2_{{1,2}} &\leq& 50^2 \quad \mbox{GeV}^2 \quad \mbox{for}\quad r\sim 50\,,\nn\\
m^2_{{1,2}} &\leq& 10 ^2\quad \mbox{GeV}^2  \quad \mbox{for}\quad  r\sim 200\,,
\eea
where $50 (10)$ GeV may be obtained only for very peculiar combination of the potential parameters. In other words we expect that also in the majority of the cases for $r$ in the range $50-200$ we will have $m^2_{{1,2}}$ very  light.

In conclusion, taking into account the SM context  and  the  potential given in \eq{A4pot},  the solution  $(e^{i \omega_1},  e^{-i \omega_1},r)v_w/\sqrt{2+r^2}$  with $r$ small, close to 1 or large give rise to very light states. Of course this does not mean that these states will be light for any value of $r$ but it is a quite strong hint that it is possible that this could be what indeed happens. As mentioned before, the addition of soft $A_4$ breaking terms to the potential may help in the cases of Goldstone bosons or very light bosons. We will discuss these terms in more detail in \cite{ABMP:Constraining2}.

\mathversion{bold}
\subsection{On the CP violation}
\mathversion{normal}
\label{secCPV}

The solutions studied in this section have an explicit complex phase $\omega_1$ in some of the vevs. One might thus wonder whether the Higgs sector in $A_4$ models gives rise to extra sources of CP violation. This CP violation can be either explicit if it appears directly at the level of the Higgs potential or implicit if it occurs due to the vevs of the scalars. In the $A_4$ Higgs scenario we are considering in this paper, neither of the two possibilities is present\footnote{This section owes to Ref.~\cite{DekensThesis} in which the question of CP violation in our class of models was first discussed in detail.}.

We first investigate whether the potential in \eq{A4pot} exhibits explicit CP violation. We find that the potential is not invariant under a ÔnaiveÕ CP transformation
\beq
\Phi_i \stackrel{CP}{\longrightarrow}\Phi_i^*\,.
\label{CPtransf}
\eeq
Under this transformation $\epsilon$ and $-\epsilon$ get interchanged in the potential in \eq{A4pot}. The expression in \eq{CPtransf} does not describe the most general CP transformation however. A more general CP transformation follows when the ÔpureÕ CP transformation in \eq{CPtransf} is combined with a Higgs basis transformation
\beq
\Phi_i \stackrel{CP}{\longrightarrow}\cU_{ij}\,\Phi_j^*\,.
\label{newCPtransf}
\eeq
Here $\cU$ is a unitary matrix in the space of the three Higgs fields. It was shown in Ref.~\cite{Branco:2005em} that the Higgs potential conserves CP explicitly if a matrix $\cU$ exists such that the ÔnewÕ CP transformation in \eq{newCPtransf} leaves the potential invariant. For the potential in \eq{A4pot} it is not hard to find such a matrix. An example is the matrix that parameterizes the interchange of the first and second Higgs fields
\beq
\cU=e^{i\alpha}\left(
        \begin{array}{ccc}
            0 & 1 & 0 \\
            1 & 0 & 0 \\
            0 & 0 & 1 \\
        \end{array}
\right)\,.
\label{UforCPcons}
\eeq
In this case, the CP transformation is defined according to
\beq
\Phi_1 \stackrel{CP}{\longrightarrow}e^{i\alpha}\Phi_2^*\,,\qquad\qquad
\Phi_2 \stackrel{CP}{\longrightarrow}e^{i\alpha}\Phi_1^*\,,\qquad\qquad
\Phi_3 \stackrel{CP}{\longrightarrow}e^{i\alpha}\Phi_3^*\,.
\eeq
We conclude that the $A_4$ invariant Higgs potential does not violate CP explicitly. 

There is still the possibility of spontaneous CP violation through the complex vacua discussed in the previous section. In Refs.~\cite{Branco:2005em,Lavoura:1994fv}, it is shown that a vacuum does not give rise to spontaneous CP violation if there is a matrix $\cU$ such that the CP transformation in \eq{newCPtransf} also leaves the vacuum invariant. In that case, the vacuum thus satisfies
\beq
\vev{\Phi}=\cU \vev{\Phi}^*\,.
\eeq
In other words, each component $v_i\, e^{i\omega_i}$ of the vector of vevs should be written as a linear combination
of the complex conjugates of the vevs $v_j\,e^{i\,\omega_j}$ with the coefficients given by $\cU_{ij}$
\beq
v_i\, e^{i\omega_i}=\cU_{ij}\,v_j\,e^{i\,\omega_j}\,.
\eeq
In the specific case under investigation, where $\cU$ has the form in \eq{UforCPcons}, this is represented by
\beq
v_1\, e^{i\omega_1}=v_2\, e^{i(\alpha-\omega_2)}\,,\qquad\qquad
v_2\, e^{i\omega_2}=v_1\, e^{i(\alpha-\omega_1)}\,,\qquad\qquad
v_3\, e^{i\omega_3}=v_3\, e^{i(\alpha-\omega_3)}\,.
\label{CondSpontCPCons}
\eeq
The first two equations are dependent: they require $v_1$ and $v_2$ to be each others complex conjugate. The third equation requires the third vev to be real. The two vacua that could lead to spontaneous CP violation, $(v\,e^{i\omega_1}, v, 0)$ and $(v\,e^{i\omega_1}, v\,e^{-i\omega_1}, r\,v)$, both satisfy the conditions in \eq{CondSpontCPCons}, for $\alpha=\omega_1$ and $\alpha=0$, respectively. As a result, they do not break CP spontaneously, notwithstanding the fact that they are inherently complex.

The criterium of conserving or violating CP depending on whether the transformation matrix $\cU$ exists, is not always a very practical one. Even if such a transformation exists, it may not be easy to find. An alternative test is in the straightforward calculation of CP-odd basis invariants that vanish if CP is conserved and that are non-zero if CP is violated (or, at least one of them is). Invariants for the potential in \eq{A4pot} and the vacua of the previous subsection were calculated in Ref.~\cite{DekensThesis}. As expected, they are all zero.

This analysis has first appeared in Ref.~\cite{ToroopThesis} and agrees with the conclusions of a recent paper in Ref.~\cite{Holthausen:2012dk}.

%
%
\section{Bounds From The Higgs Phenomenology}
\label{sec:bounds}

In this section we analyse the phenomenology corresponding to the different vacua discussed above: unitarity, $Z$ and $W^\pm$ decays and oblique parameters. In this way we manage to constrain the parameter space and, in some cases, to rule out the studied vacuum configuration.

\subsection{Unitarity}

In this section we account for the tree level unitarity constraints coming from the additional scalars present in the theory. We examine the partial wave unitarity for the neutral two-particle amplitudes for $s\gg M_W^2,M_Z^2$. We can use the equivalence theorem, so that we can compute the amplitudes using only the scalar potential described in eq. (\ref{A4pot}). In the regime of large energies, the only relevant contributions are the quartic couplings in the scalar potential \cite{Veltman:Unitarity,LQT:Unitarity1,LQT:Unitarity2,CDFG:Unitarity} and then we can write the $J=0$ partial wave amplitude $a_0$ in terms of the tree level amplitude $T$ as
\beq
a_0(s)\equiv\dfrac{1}{32\pi}\int_{-1}^{1}\di\!\cos\theta\; T(s)=\dfrac{1}{16\pi} F[\lambda_i]\,,
\eeq
where $F$ represents a function of the $\lambda_i$ couplings. Using for simplicity the notation
\beq
\Phi_a =
\bmat
w_a^+ \\[2mm]
\dfrac{v_a e^{i\omega_a}+h^0_a+ i z_a}{\sqrt2}
\emat\;,
\eeq
we can write the 30 neutral two-particle channels as follows:
\beq
w_a^+w_b^-\,,\;\dfrac{z_az_b}{\sqrt2}\,,\;\dfrac{h^0_ah^0_b}{\sqrt2}\,,\;h^0_az_b\,.
\eeq
Once written down the full scattering matrix $a_0$, we find a block diagonal structure. The first $12\times12$ block concerns the channels
\beq
w_1^+  w_1^-\,,\; w_2^+ w_2^-\,,\; w_3^+ w_3^-\,,\; \dfrac{z_1 z_1}{\sqrt2}\,,\; \dfrac{z_2 z_2}{\sqrt2}\,,\; \dfrac{z_3 z_3}{\sqrt2}\,,\; \dfrac{h^0_1 h^0_1}{\sqrt2}\,,; \dfrac{h^0_2, h^0_2}{\sqrt2}\,,\; \dfrac{h^0_3, h^0_ 3}{\sqrt2}\,,\; h^0_1z_1\,,\; h^0_2z_2\,,\; h^0_3 z_3\,,\nn
\eeq
while the other three $6\times6$ blocks are related to the channels
\beq
w_a^+  w_b^-\,,\; w_b^+ w_a^-\,,\; h^0_a z_b\,,\; h^0_b z_a\,,\; z_a z_b\,,\; h^0_a h^0_b\,,\nn
\eeq
once we specify the labels $(a,\,b)$ as $(1,\,2)$, $(1,\,3)$ and $(2,\,3)$. Notice that up this point the analysis is completely general and is valid for all the vacua presented. We specify the vacuum configuration, expressing the quartic couplings $\lambda_i$ in terms of the masses of the scalars. Afterwards, putting the constraint that the largest eigenvalues of the scattering matrix $a_0$ is in modulus less than 1, we find upper bounds on the scalar masses which we use in our numerical analysis.

\mathversion{bold}
\subsection{$Z$ And $W^\pm$ Decays}
\mathversion{normal}

From an experimental point of view gauge bosons decays into scalar particles are detected by looking at fermionic channels, such as  for example $Z\to h A\to 4 f$    in the 2HDM, or $Z$ decays into  partial or total missing energy in a generic new physics scenario. From this point of view gauge bosons  decays  bound  the Higgs sector in an  extremely model dependent way. However since in the SM the $Z$ and the $W^\pm$  decays into 2 fermions, 4 fermions or \emph{all}  have  been precisely been calculated and measured, we may focus on the decays $Z ,W^\pm\rightarrow\, all$.  Doing this we overestimate the allowed regions in the parameter space, but we  have a first and model independent cut arising by the gauge bosons decay.  Once we will pass to a model dependent analysis the region may only be restricted, not enlarged. Furthermore, defining the contribution from new physics as $\Delta\Gamma$, since
\beq
\Delta \Gamma^{2f}_{Z,W^\pm}\sim \Delta \Gamma^{4f}_{Z,W^\pm}\sim  \Delta \Gamma^{all}_{Z,W^\pm}\ll \Gamma_{Z,W^\pm}\,,
\eeq
we expect the error we commit  being quite small.

From LEP data  we have
\beq
\Gamma_{Z, W^\pm}^\textrm{exp}= \Gamma_{Z, W^\pm}^\textrm{SM} + \Delta\Gamma_{Z,W^\pm}
\eeq
with $\Delta \Gamma_{Z}\sim 0.0023$ GeV and $\Delta \Gamma_{W^\pm}\sim 0.042$ GeV \cite{PDG2010}. Therefore we may calculate the width
\beq
\ba{ccl}
Z&\to&   h_i h_j\,,\\[2mm]
W^+&\to& H^+_i h_j \,.
\ea
\ee
for the different multi Higgs (MH)  vacuum configuration studied and select the points that satisfy
\beq
\Gamma^{MH}_{Z,W^\pm}\leq \Delta \Gamma_{Z,W^\pm}\,.
\eeq
Here we have indicated the generic $Z\to h_i h_j$ referring to our notation introduced in section \ref{sec:potential}. Clearly when CP is conserved the $h_i$ have defined CP and only couplings to CP odd states are allowed.  

In the vacuum  analysis  we did we have seen that in  few situations we have extra massless or very light particles. For those cases the gauge bosons decays put  strong bounds.  For what concerns the $Z$ decays we have
\beq
\label{eqZdecay}
\begin{cases}
k_Z \leq \Delta \Gamma_Z\dfrac{16 \pi}{m_Z} \dfrac{4 c_W^2}{g^2}&\qquad \mbox{ if both particles $h_i$ and $h_j$ are massless}\,,\\[2mm]
k_Z \left(1 - \dfrac{m_{h_i}^2}{m_Z^2}\right)^3\leq\Delta \Gamma_Z \dfrac{ 16 \pi}{m_Z} \dfrac{4 c_W^2}{g^2}&\qquad \mbox{ if $h_j$ is masslees and  $0<m_{h_i}^2 <m_Z^2$}\,,\\[2mm]
k_Z \left(1 - \dfrac{m_{h_i}^2+m_{h_j}^2}{m_Z^2}\right)^3\leq\Delta \Gamma_Z \dfrac{ 16 \pi}{m_Z} \dfrac{4 c_W^2}{g^2}&\qquad \mbox{ if $h_i,h_j\neq 0$  and  $0<m_{h_i}^2+m_{h_j}^2 <m_Z^2$}\,.
\end{cases}
\eeq
where $g$ is the $SU(2)$ gauge coupling, $c_W$ the cosine of the Weinberg angle $\theta_W$ and the parameter $k_Z$ is given by
\beq
k_Z=\left(-U^T_{a b} U^T_{(a + 3) c} + U^T_{(a+3) b} U^T_{a  c}\right)^2\,,
\eeq
with $U$  defined in \eq{defU}.

Similarly for the $W^\pm$ decays we have
\beq
\label{eqWdecay}
\begin{array}{lll}
k_W \left(1 - \dfrac{m_{C_i}^2}{m_W^2}\right)^3\leq\Delta \Gamma_W \dfrac{ 16 \pi}{m_W} \dfrac{4 c}{g^2}&\quad& \mbox{ if $h_j$ is massless and $m_{C_i}^2 <m_W^2$}
\end{array}
\eeq
where, in analogy to the $Z$ decay, the parameter $k_W$ is given by
\beq
k_W= \left|S^\dag_{a b} U^T_{a c} \right|^2+ \left|S^\dag_{(a+3) b} U^T_{(a+3) c} \right|^2\,,
\eeq
with $S$ defined in \eq{defS}.

\subsection{Large Mass Higgs Decay}

Electroweak data analysis considering the data from LEP2 \cite{LEP2} and Tevatron \cite{Tevatron} put an upper bound on the mass of the SM Higgs of $194$ GeV at $99\%$ CL \cite{PDG2010}. In a MH scenario this bound may be roughly translated in the upper bound for the lightest scalar mass, $m_{h_1}$. For large values of the SM Higgs mass, $m_h\geq 2 m_W$,  the main channel decay is $h\to W^+ W^-$ and the upper bound is completely model independent.  Let us indicate as $\Gamma^{SM}_{WW}(194)$ the branching ratio of the SM Higgs into two $W^\pm$ at a mass of $194$ GeV.

In a MH model the lightest Higgs boson couples to the gauge bosons with a coupling that is
\beq
\ba{rcl}
g_{h_1 ZZ}&=& \beta \,g_{hZZ}^{SM}\,, \\[2mm]
g_{h_1 WW}&=& \beta\, g_{hWW}^{SM}\,,
\ea
\ee
with $\beta\leq 1$. In our case for example $\beta$ is given by
\beq
f_a (\cos \omega_a\, U^T_{a 1}+ \sin \omega_j\, U^T_{(a+3) 1})\,,
\eeq
with $f_a=v_a/v_w$ and $\omega_a$ the corresponding CP phase. Taking into account that $h_1$ is less produced then the SM Higgs and that its $\Gamma^{MH}_{ WW}(m_{h_1})$ is reduced with respect to the SM one,
\beq
\Gamma^{MH}_{ WW}(m_{h_1})\sim |\beta|^4 \Gamma^{SM}_{WW}(m_{h_1})\leq \Gamma^{SM}_{WW}(194)\,,
\eeq
we can roughly constrain the upper bound for masses $m_{h_1}\geq 194$ GeV.

\subsection{Constraints By Oblique Corrections}
\label{sec:stu}
The consistence  of a  MH model   has to be checked also   by means of the oblique corrections.
These corrections can be classified \cite{HT:TSUparameters,PT:TSUparameters1,GR:TSUparameters,DEH:TSUparameters,PT:TSUparameters2} by means of three parameters, namely  $TSU$, that  maybe written in terms of the physical gauge boson vacuum polarizations as \cite{MBL:TSUparameters}
\beq
\label{STU2}
\hspace{-3mm}
\ba{rcl}
T & = & \dfrac{4 \pi }{e^2 c_W^2 m_Z^2} \left[ A_{WW}(0)- c_W^2 A_{ZZ} (0) \right] \,, \\[3mm]
S & = &16 \pi \dfrac{ s_W^2 c_W^2}{e^2}  \left[  \dfrac{A_{ZZ}(m_Z^2)- A_{ZZ}(0) }{m_Z^2}- A'_{\gamma \gamma}(0) -\dfrac{ (c_W^2-s_W^2) }{c_W s_W} A_{\gamma Z}'(0) \right] \,, \\[3mm]
U &=& -16 \pi \dfrac{s_W^2}{e^2} \left[ \dfrac{A_{WW}(m_W^2)- A_{WW}(0) }{m_W^2}- c_W^2 \dfrac{A_{ZZ}(m_Z^2)- A_{ZZ}(0) }{m_Z^2}-s_W^2 A_{\gamma\gamma}' (0) - 2 s_W c_W A_{\gamma Z}' (0) \right]\,,
\ea
\ee
where  $s_W,c_W$ are sine and cosine of $\theta_W$ and $e$ is the electric charge.
EW precision measurements severely constrain the possible values of the three parameters $T$, $S$ and $U$.
 In the SM  assuming $m^2_h>m^2_Z$ we have
\be
\label{EPSM}
\ba{rcl}
T^{SM}_{h}&\sim& -\dfrac{3 }{16 \pi c_W^2} \log \dfrac{m^2_h}{m^2_Z}\,,\\[2mm]
S^{SM}_{h}&\sim& \dfrac{1}{12 \pi}\log \dfrac{m^2_h}{m^2_Z}\,,\\[2mm]
U^{SM}_h&\sim&0\,.
\ea
\ee
For a Higgs boson mass of $m_h=$ 117 GeV (and in brackets the difference assuming instead $m_h=300$ GeV), the data allow \cite{PDG2010}
\beq
\label{EPEXP}
\ba{rcl}
S^{exp} &=& 0.10  \pm 0.10 (-0.08)  \\[2mm]
T^{exp} &=& 0.03   \pm 0.11(+0.09) \\[2mm]
U^{exp} &= & 0.06\pm   0.10(+0.01)\,.
\ea
\ee
The constraints in \eq{EPEXP} must be rescaled not only for the different values of the Higgs boson mass but also for a different scalar or  fermion field content: for example, if we assume to have a MH scenario this gives a contribution $T^{\textrm{MH}}$   to the T-parameter  and  we need
\be
T^{\textrm{NSS}} - T^{\textrm{SM}}_{h}= T^{exp}\,.
\ee
A detailed analysis on the $TSU$ in a MH model has been presented in  \cite{GLOO:TSUparameters1,GLOO:TSUparameters2} where  all the details are carefully explained.  However the resulting formulae are valid only for scalar masses larger or comparable to $m_Z$. Since  this is not the case for a generic MH model and particularly for the configurations studied so far, where we have a redundant number of massless or extremely light particles, we improved their results, getting full formulae valid for any value of the  scalar masses (see the appendix A for details).

%
%
\section{Results}
 \label{sec:results}

We have performed a numerical  analysis for all vacuum configurations considered, neglecting the alignment $(v,v,0)$  since in this case there are tachyonic states. Our aim was to find  a region in the parameter space where all the Higgs constraints were satisfied for each configuration considered.  We have analysed the points generated  through subsequent constraints, from the weaker one to the stronger according to
  \begin{itemize}
  \item points Y: true minima --all the squared masses positive-- (yellow points in the figures);
  \item points B: unitarity bound (blue points);
  \item points G: $Z$ and $W^\pm$ decays (green points);
  \item points R: $TSU$ parameters (red points).
    \end{itemize}
The ratios $B/Y$, $G/B$, $R/G$ may be used to indicate which is the  stronger constraint for each allowed minima.
For almost each case we have compared the masses of the two lightest neutral states --except for the alignment studied in sec.~\ref{seccomplecvevscasei} where we have only one massive neutral state-- and the mass of the lightest neutral scalar versus the mass of the lightest charged one.
Then we have plotted the $TS$ oblique  parameters for all the green points to check that $T$ is the most constrained one --for this reason we have not inserted the plots concerning $U$.

On the contrary for the alignment $(v e^{i \omega_1},v e^{-i \omega_1},r v)$  we have personalized the plots for reasons that will be clear in the following.

Notice that in all the following discussion, we refer as $m_1$ ($m_2$) to the (next-to-the-) lightest neutral state and as $m_{ch_1}$ as the lightest charged mass state.

\subsection{Solutions with real vevs}

\mathversion{bold}
\subsubsection{The Alignment $(v,v,v)$}
\mathversion{normal}
\label{ressecCP1}

\begin{figure}[ht!]
\centering
\includegraphics[width=7cm]{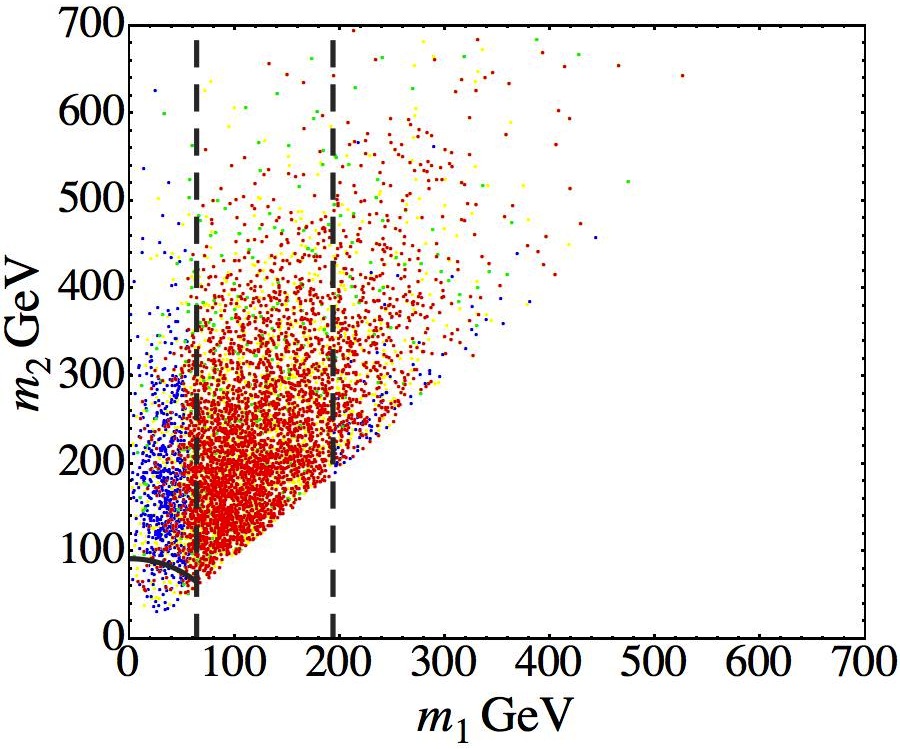}
\includegraphics[width=7cm]{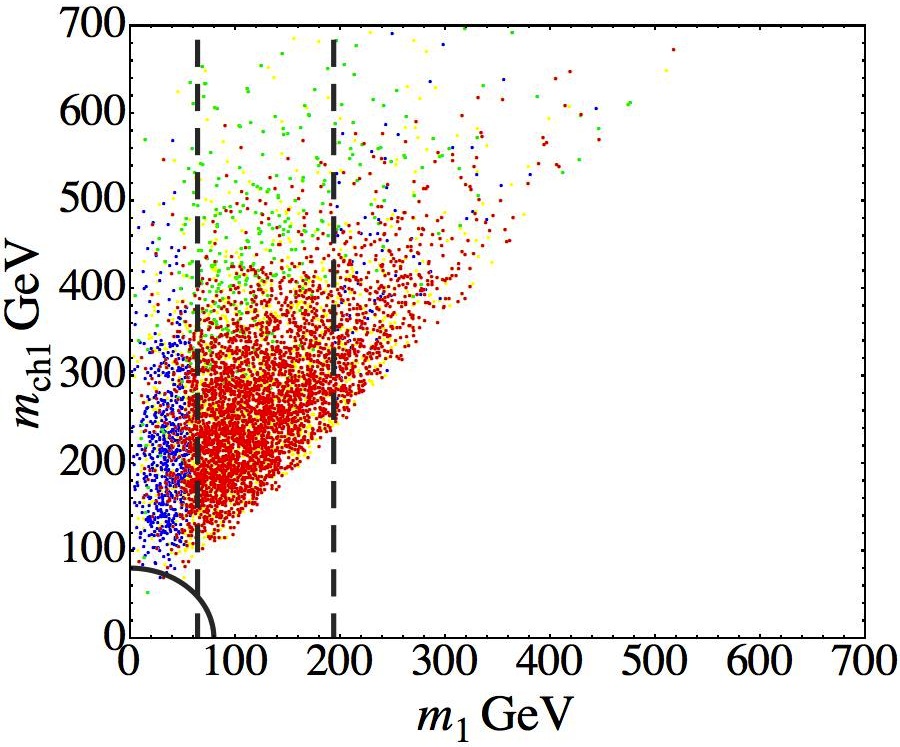}\\
\includegraphics[width=7cm]{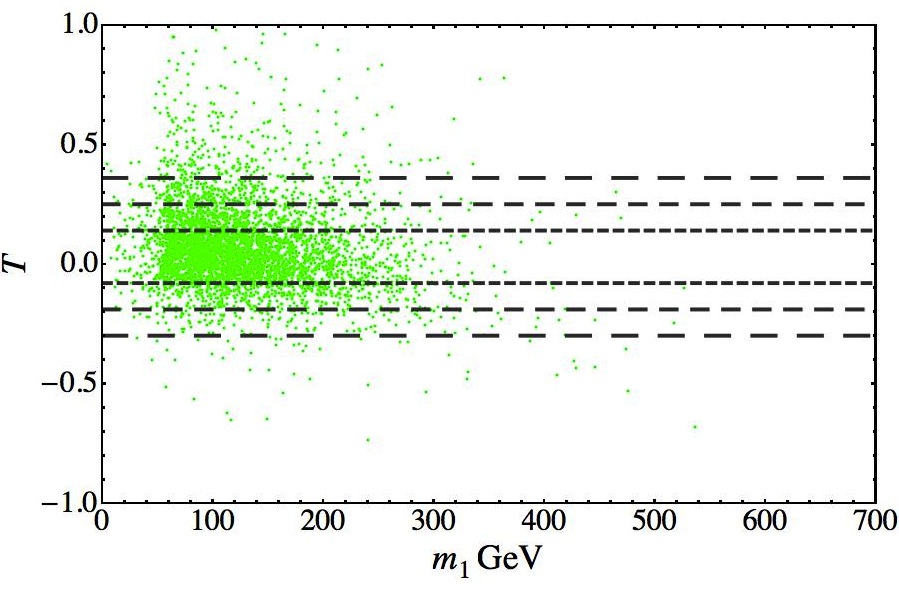}
\includegraphics[width=7cm]{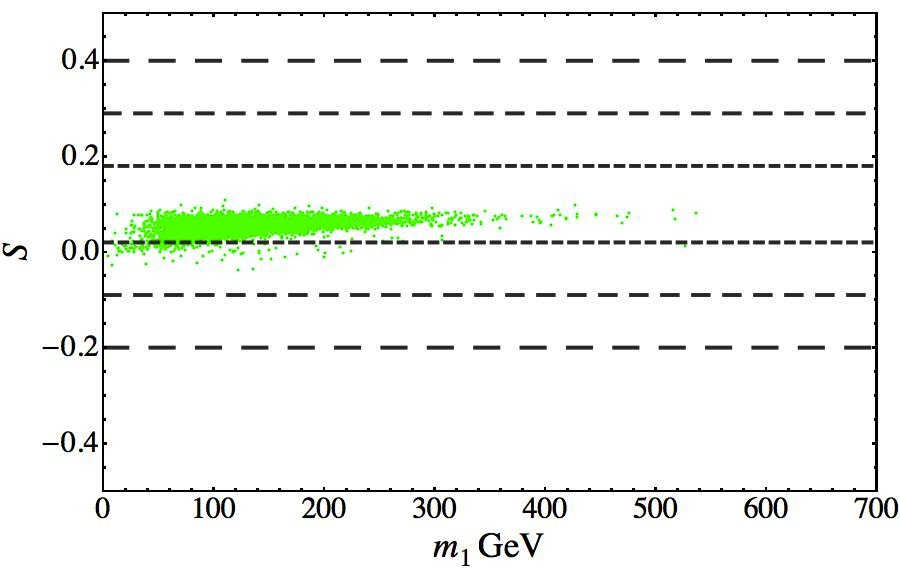}
\caption{Alignment $(v,v,v)$: \it the upper panels show the lightest neutral mass $m_1$ versus the second lightest neutral mass $m_2$ and the  lightest  charged one $m_{ch_1}$ respectively. The gray arc delimits the region below which the $Z$ ( $W$) decay channel opens. On the left plot the arc is only of $45^\circ$ because $m_2\geq m_1$. For points below the arc the $ Z$ ( $W$) decay may happens. The points allowed stretch in the region close to the border because of the conditions of  \eq{eqZdecay}.  The dashed vertical lines indicates the approximated cuts that occur at $m_1\sim m_Z/\sqrt{2}$ and $m_1\sim 194 $ GeV according to case $2)$ and case $1)$ respectively as explained in the text. The down panels show the contributions to $T$ and $S$ for the G points. The  gray dashed lines indicate the experimental values at $3$, $2$, $1\sigma$ level --long,normal,short dashing respectively. The $T$ parameter turns out to be the most constraining one.}
\label{fig.vvv}
\end{figure}

In sec.  \ref{secvvv} we have redefined the initial 3 doublets  in term of the $Z_3$ surviving symmetry representation: $\bf1$, $\bf1'$, $\bf1''$. One combination corresponds to a  $Z_3$ singlet doublet, that behaves like the SM Higgs: it develops a non-vanishing vev, gives rise to a CP even state which we call $h_1$ and to the three GBs eaten by the gauge bosons. The others two doublets, $\varphi'$ and $\varphi''$, are inert.  From these informations we may already figure out what we expect by the numerical scan:
\begin{itemize}
\item[1)] when $m_{h_1}$ is the smallest mass, $h_1$ is the lightest state and corresponds to the SM-like Higgs. As a result, the usual SM mass upper bound applies. On the contrary as long as we do not consider its coupling with the fermions we do not have a  model independent  lower mass bound. This is due to a combined effect of the CP and $Z_3$ symmetries: $h_1$ is CP even  and singlet under $Z_3$, but couplings like $Z h_1\varphi^{\prime0}$, $Z h_1\varphi^{\prime\prime0}$, $W^-h_1\varphi^{\prime1}$  or $W^-h_1\varphi^{\prime\prime1}$ are forbidden because of $Z_3$ and then gauge boson decays cannot constrain the lower mass of $h_1$.
\item[2)] When $\varphi^{\prime0}$ ($\varphi^{\prime\prime0}$) is the lightest state, we do not have an upper bound on this state because the couplings $\varphi^{\prime0} W^+ W^-$ ($\varphi^{\prime\prime0} W^+ W^-$) is absent. On the contrary we may have a lower bound because couplings like $ Z\varphi^{\prime0}\varphi^{\prime\prime0}$ and $W^- \varphi^{\prime0} \varphi^{\prime\prime1} $ are allowed.
\end{itemize}
Combining the two situations sketched in points $1)$ and $2)$, we expect neither lower nor upper bounds for the lightest Higgs mass: according to which of the two cases is most favored, we may expect a denser vertical region around $m_1\sim m_Z/\sqrt{2}$ when the $Z$ decay channel closes according to \eq{eqZdecay} --case $2)$ more favored-- or a denser vertical line around $m_1\sim 194$ GeV, if the large Higgs mass decay constrain applies --case $1)$ more favored. Indeed by looking at fig.~\ref{fig.vvv} we see that we may find R (allowed) points for very tiny $m_1$ masses and up to $\sim 500$ GeV when the unitarity bound starts to show its effect. However by looking at the crowded points in  fig.~\ref{fig.vvv} it seems that case $2)$ is slightly preferred with respect to case $1)$. Finally  for the G points --those that satisfy the minimum, unitarity and decays conditions-- we have compared the contributions to the oblique parameters $T$ and $S$ to see which of the two is more constraining. It turns out to be $T$, while we have not reported $U$ because its behavior is very similar to $S$.


 \mathversion{bold}
\subsubsection{The Alignment $(v,0,0)$}
\mathversion{normal}
\label{ressecCP2}

\begin{figure}[ht!]
\centering
\includegraphics[width=7cm]{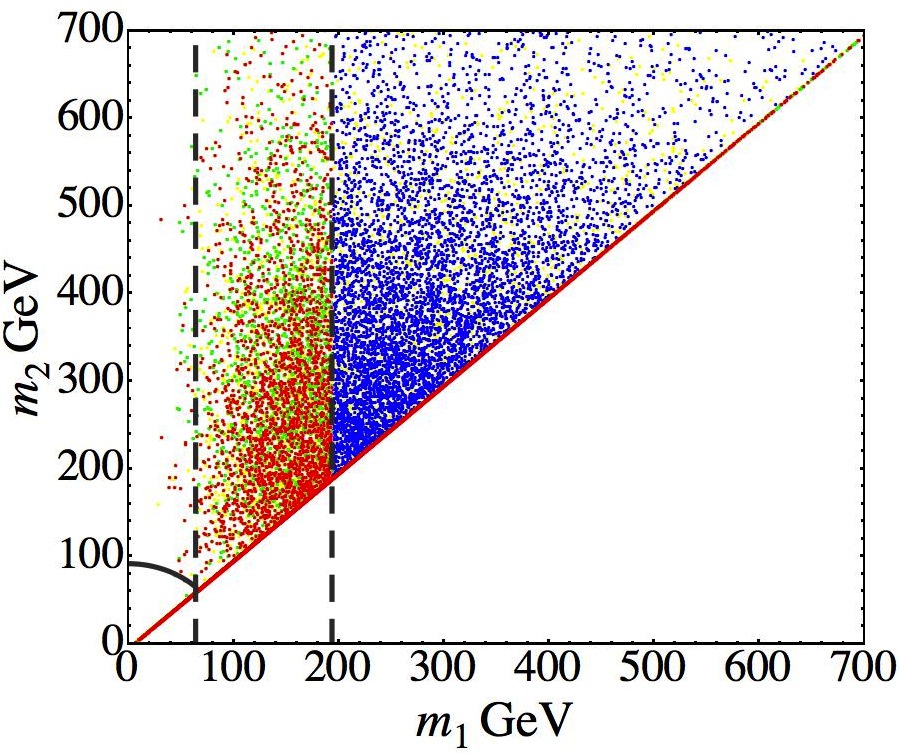}
\includegraphics[width=7cm]{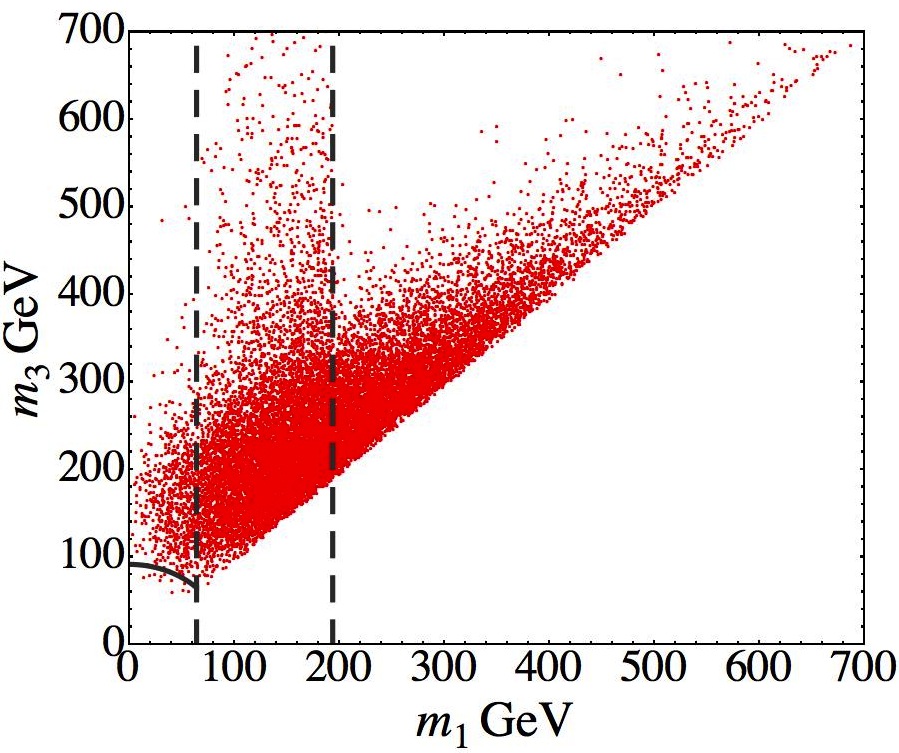}\\
\includegraphics[width=7cm]{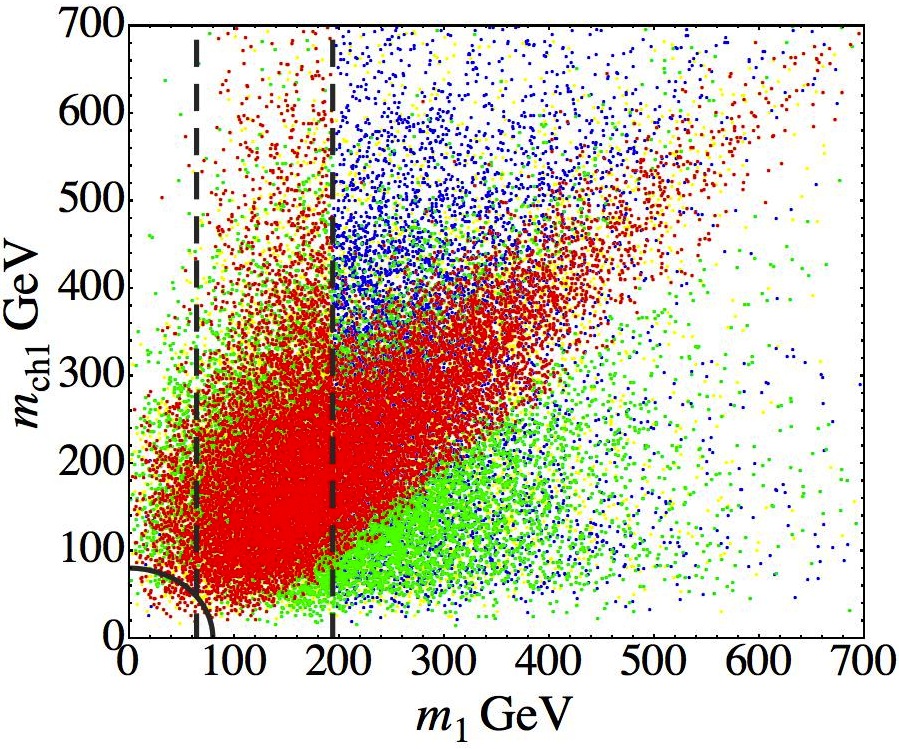}\\
\includegraphics[width=7cm]{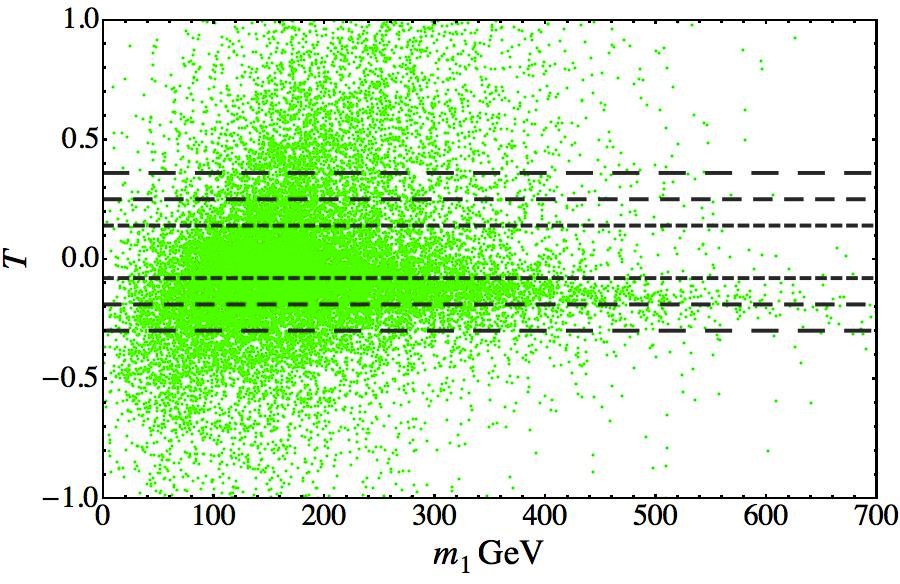}
\includegraphics[width=7cm]{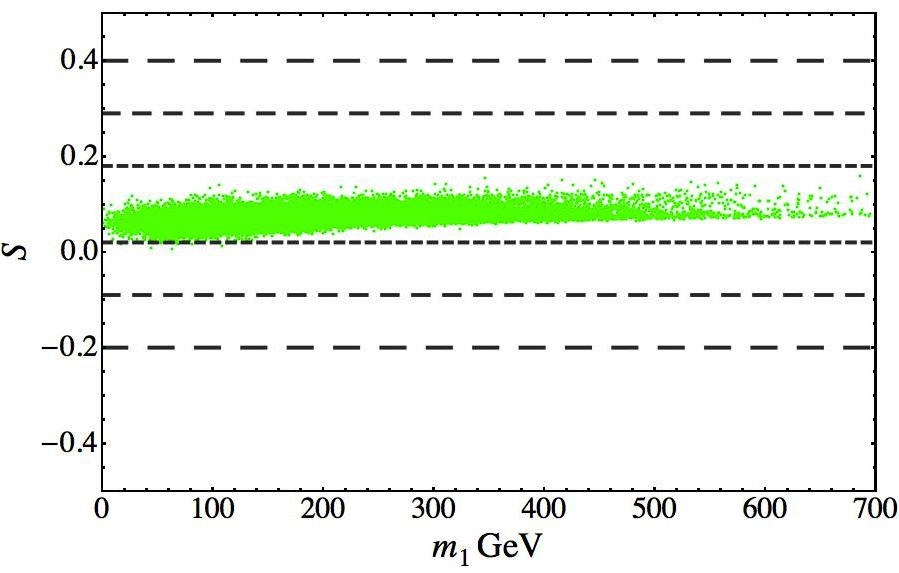}
\caption{{Alignment $(v,0,0)$}: \it the upper panels show $m_1$ versus $m_2$ (on the left) and third lightest $m_3$ (on the right). For the latter we reported only the R points. The central panel shows $m_1$ versus $m_{ch_1}$. The gray arc delimits the region below which the $Z$ ($W$) decay channel opens while the second dashed vertical one the SM-Higgs mass upper bound at $194$ GeV.  The first dashed vertical line at $m_1=m_Z/\sqrt{2}$ is reported to help a comparison with  the $Z_3$ preserving case. On the first two plots  the arc is only of 45 degrees because $m_{2,3}\geq m_1$.  The down panels show the contributions to $T$ and $S$ for the G points. The $T$ parameter turns out to be the most constraining one.}
\label{fig.v00}
\end{figure}

For what concerns the second natural $A_4$ minimum, the $Z_2$ preserving one, things slightly change with respect to the $Z_3$ surviving case. By sec. \ref{secv00} we know that as for the $Z_3$ case we have a SM-like doublet, $Z_2$ even, that develops the vev, gives rise to a CP even neutral state, $h_1$, and to the GBs eaten by the gauge bosons. However contrary to the $Z_3$ case, in the $Z_2$ minima we have 4 $Z_2$ odd states, 2 CP even labelled $h_{2,3}$  and  2 CP odd labelled $h_{4,5}$. Moreover the 2 CP even (odd) are degenerate. As done in sec. \ref{ressecCP1} we may sketch what we expect from the numerical analysis:
\begin{itemize}
\item[1)] when $h_1$, the $Z_2$ even SM-like Higgs, is the lightest we expect  the SM Higgs upper bound but no lower bound because the interactions $Z h_1 h_{4,5}$ are forbidden by the $Z_2$ symmetry;
\item[2)] when the two lightest are the $Z_2$ odd degenerate states $h_{2,3}$ --CP even-- or $h_{4,5}$ --CP odd-- we expect no upper bound. Moreover since they are degenerate we do not expect lower bound too. On the contrary we expect that $Z$ and $W$ decays constrain the third lightest neutral Higgs mass and that of the  charged ones.
\end{itemize}
By looking at fig. \ref{fig.v00} we see that indeed we have a large amount of points for which $m_1=m_2$ for values from 0 up to $700$ GeV, thus reflecting case $2)$. Then the points corresponding to case $1)$ have a sharp cut at $m_1=194$ GeV, that rejects many blue points, i.e. those satisfying the unitarity constrain but not the decays one.  We have reported also $m_1$ versus $m_3$ to check that indeed, when $m_1\to0$, $m_3$ is bounded by $m_Z$ as  we expected. Our intuitions are also confirmed by the plot $m_1-m_{ch_1}$. As for the $Z_3$ preserving case the most constraining oblique parameter is $T$.


\mathversion{bold}
\subsubsection{The Alignment  $(v_1,v_2,v_3)$ with $\epsilon=0$, $\lambda_3+\lambda_4+\lambda_5=0$ }
\mathversion{normal}
\label{ressecCP3}

In this case we do not have any surviving symmetry which forbid  some couplings. However from sec. \ref{realvevssub} we know that the conditions $\epsilon=0$, $\lambda_3+\lambda_4+\lambda_5=0$ give rise to two extra massless CP even particles. Therefore we  expect that
\begin{itemize}
\item[1)] when the lightest massive state is CP odd, then its mass is bounded by the $Z$ decay through \eq{eqZdecay};
\item[2)] when the lightest massive state is CP even, then its mass  could reach smaller values since the $Z$ decay bound would constrain the combination of its mass with the lightest CP odd state mass.
\end{itemize}
Moreover in both cases we expect  the mass of the lightest charged scalar bounded by $W$ decay, according to \eq{eqWdecay}, due to its coupling with $W$ and the massless particles.

\begin{figure}[ht!]
\centering
\includegraphics[width=7cm]{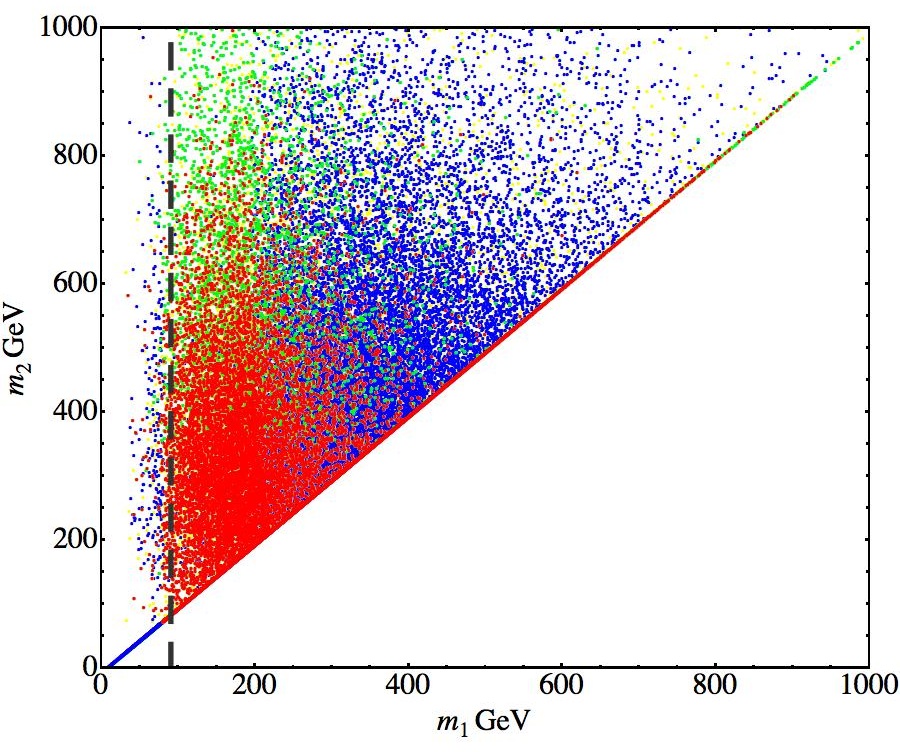}
\includegraphics[width=7cm]{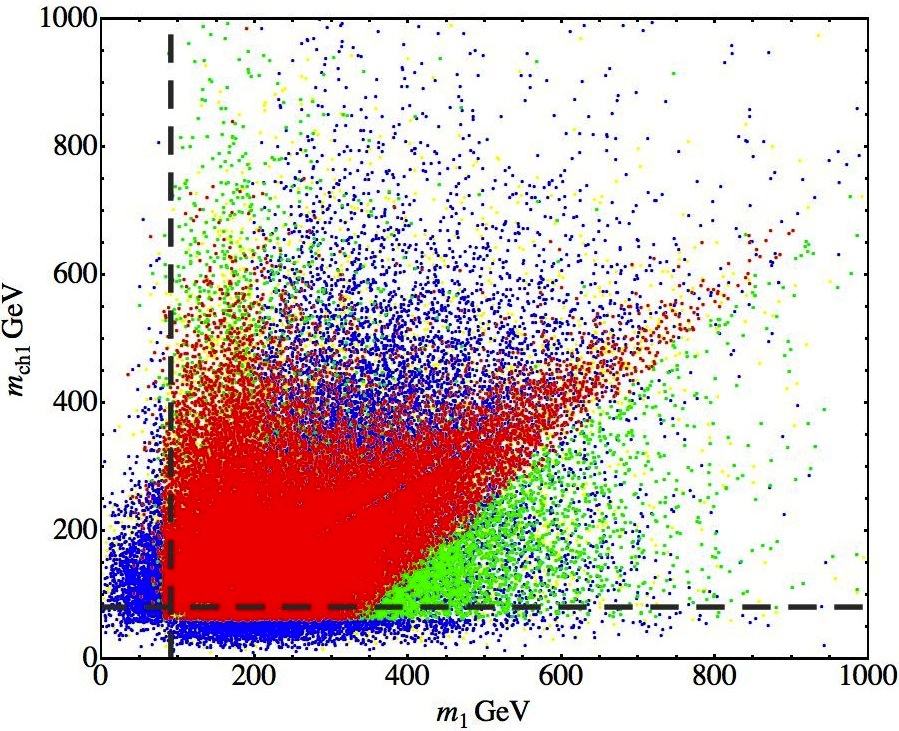}\\
\includegraphics[width=7cm]{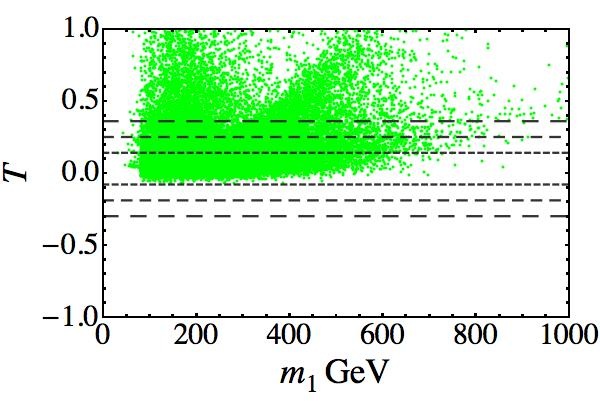}
\includegraphics[width=7cm]{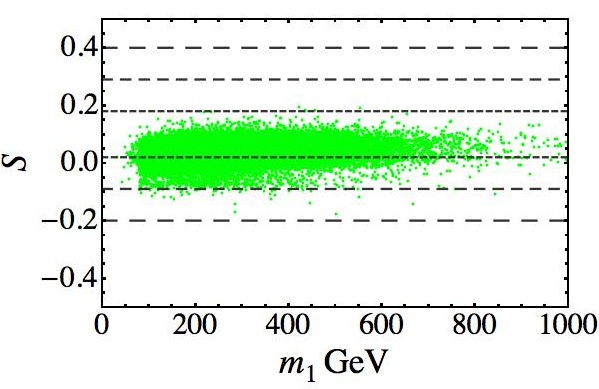}
\caption{Alignment $(v_1,v_2,v_3)$: \it the upper panels show $m_1$ versus $m_2$ and $m_{ch_1}$ respectively.  The dashed lines at $m_1=m_Z$ (vertical) and $m_{ch1}=m_W$ (horizontal) delimit the region below which the $Z$ and $W$  decay channels open respectively. The allowed points concentrate close to the borders according  to eqs.~\ref{eqZdecay}-\ref{eqWdecay}. The down panels show the contributions to $T$ and $S$ for the G points. The $T$ parameter turns out to be the most constraining one.}
\label{fig.v1v2v3}
\end{figure}

By fig.~\ref{fig.v1v2v3} we see that it seems that case $2)$ happens very rarely because the cut at $m_1\sim m_Z$ is in evidence. As for the $Z_3$ and $Z_2$ preserving minima the $T$ parameter is the most constraining one.

 \subsection{Solutions with complex vevs}
 \mathversion{bold}
\subsubsection{The Alignment $(v e^{i \omega_1}, v,0)$}
\mathversion{normal}
\label{ressecnoCP1}

\begin{figure}[ht!]
\centering
\includegraphics[width=7cm]{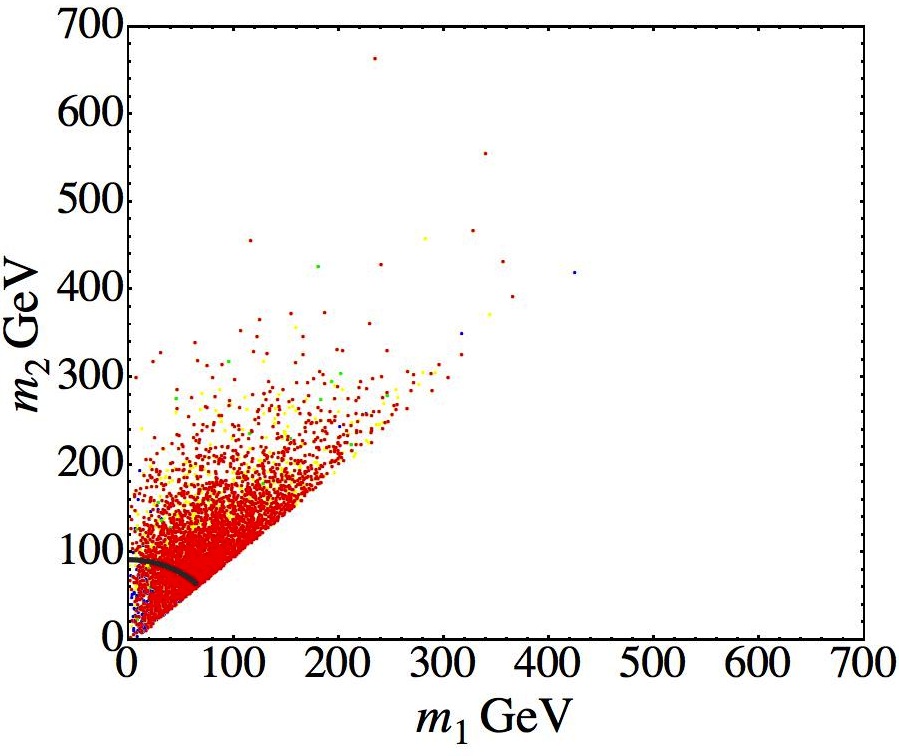}
\includegraphics[width=7cm]{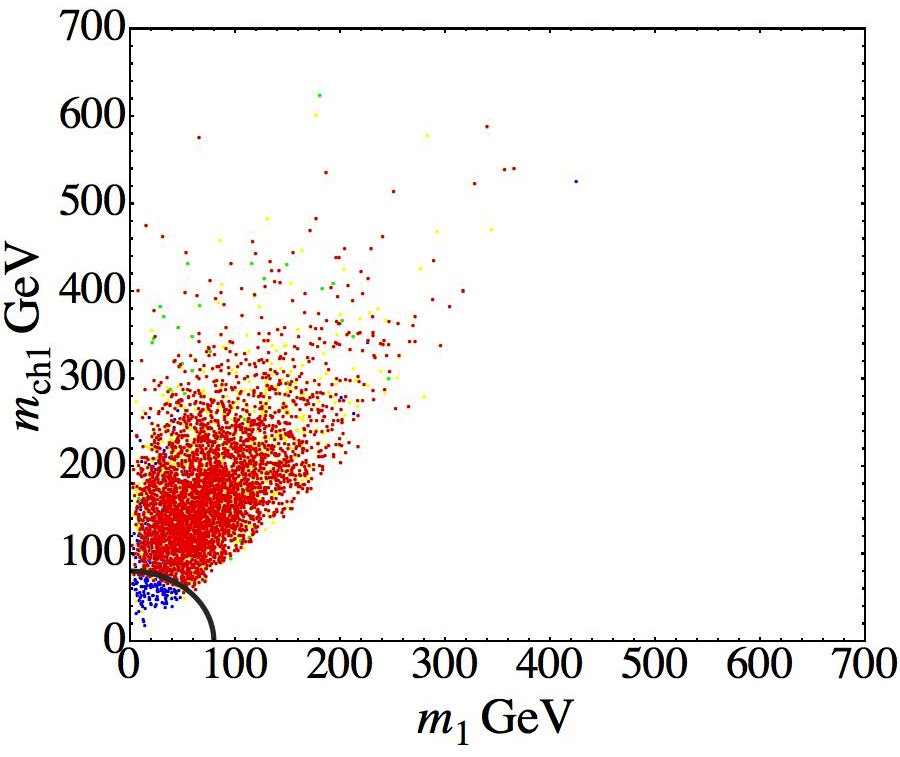}\\
\includegraphics[width=7cm]{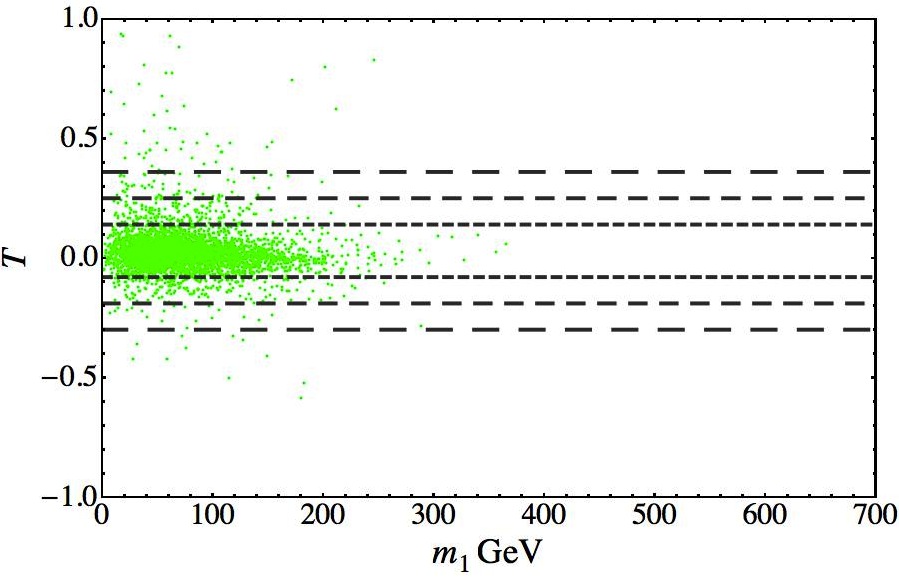}
\includegraphics[width=7cm]{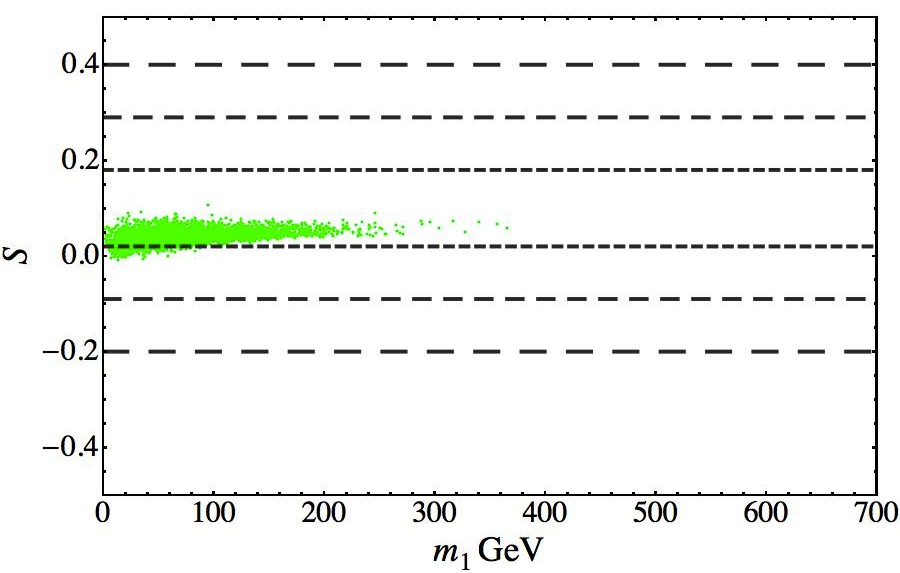}
\caption{Alignment $(v e^{i \omega_1},v,0)$: \it as in the previous figure the upper panels show $m_1$ versus $m_2$ and $m_{ch_1}$ respectively. In the plot on the right, the effect of the $W$ decay constraint on $m_1^2+m_{ch_1}^2$ is clear by looking at the B points. The down panels show the contributions to $T$ and $S$ for the G points. The $T$ parameter turns out to be the most constraining one.}
\label{fig.vvomega0}
\end{figure}

As for the vacuum alignment $(v_1,v_2,v_3)$ commented in sec~\ref{ressecCP3} the alignment $(v e^{i \omega_1}, v,0)$ does not preserve any $A_4$ subgroup. Since the two lightest Higgses might have the same CP eigenvalue, the $Z$ boson does not decay into them and no lower bound on $m_1$ and $m_2$ can be recovered in fig. \ref{fig.vvomega0}. On the other hand, the $W$ boson decay gives a lower bound on the quantity $m_1^2+m_{ch_1}^2$. Regarding the upper bound on the lightest neutral mass state we do not expect any clear cut, because we may not identify a SM-like Higgs.

\mathversion{bold}
\subsubsection{The Alignment  $(v e^{i \omega_1}, v e^{-i \omega_1},r v)$ case $i)$}
\mathversion{normal}
\label{resseccomplexvev2}

\begin{figure}[ht!]
\centering
\includegraphics[width=7cm]{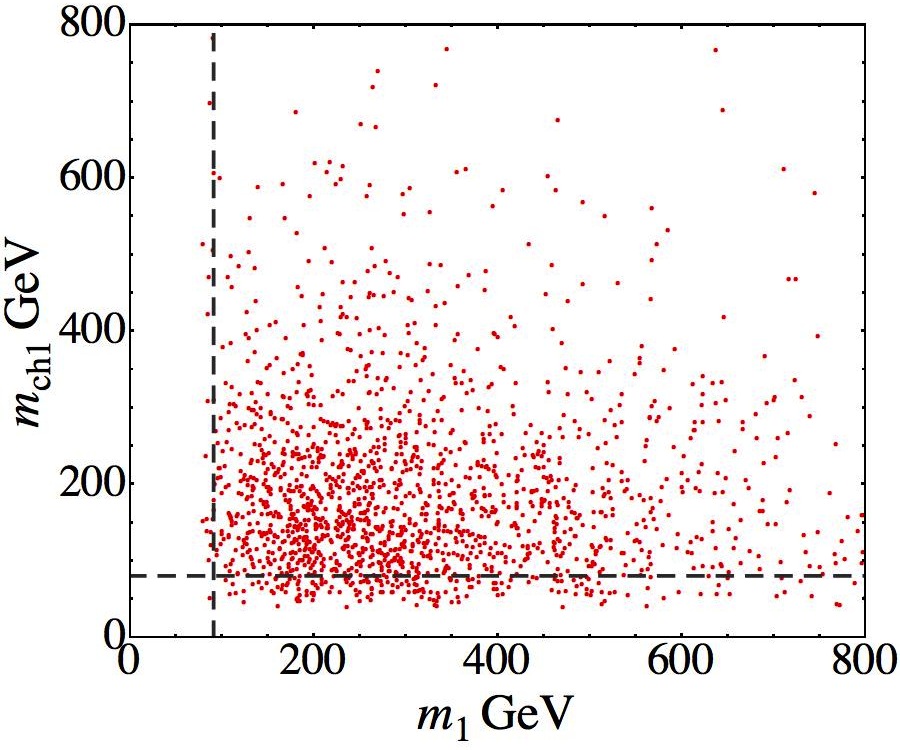}\\
\includegraphics[width=7cm]{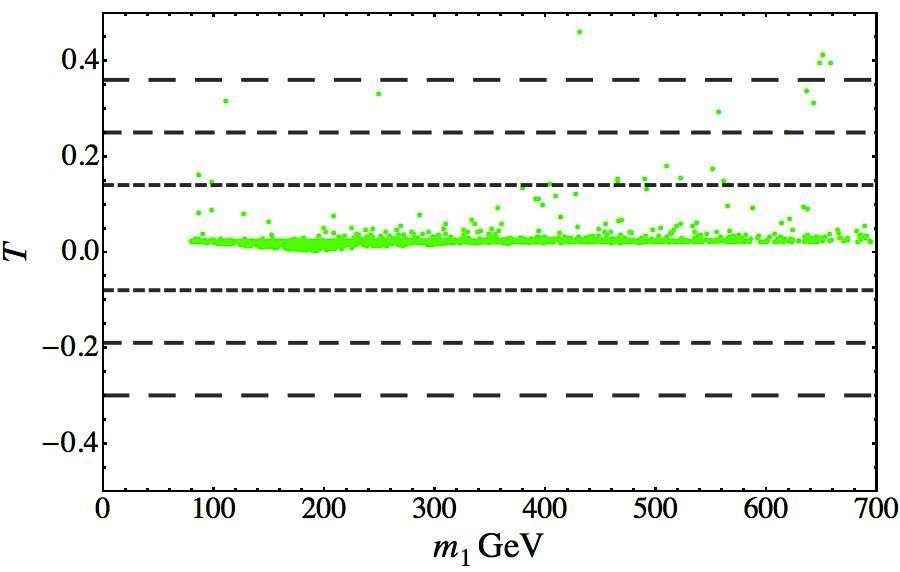}
\includegraphics[width=7cm]{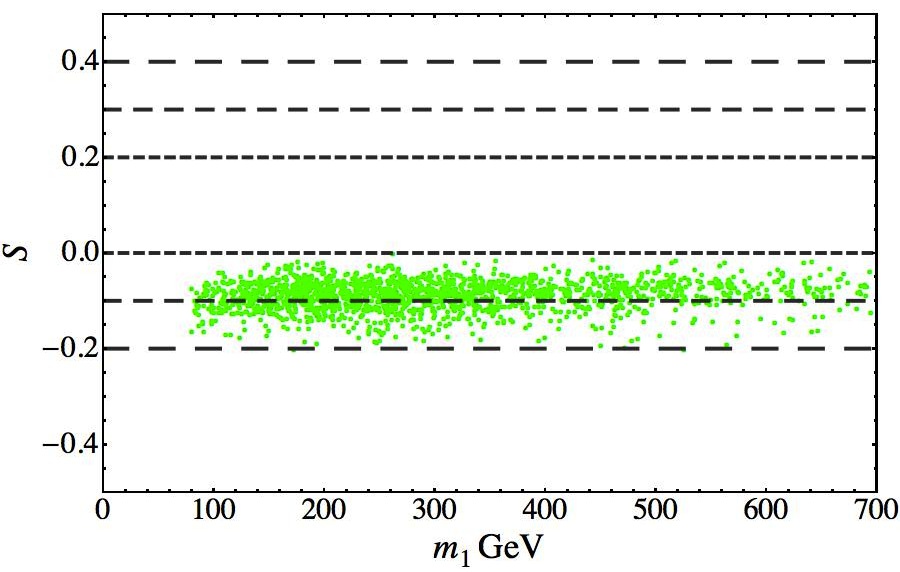}
\caption{Alignment $(v e^{i \omega_1}, v e^{-i \omega_1},r v)$ case $i)$: \it  the upper panel show $m_1$ versus $m_{ch_1}$. Only the R points are reported. The down panels show the contributions to $T$ and $S$ for the G points. For this specific case the $TSU$ oblique parameter constrain is irrelevant compared to the decay one.}
\label{fig.vvuomega1}
\end{figure}

In sec.~\ref{seccomplecvevscasei} we have seen that the alignment  $(v e^{i \omega_1}, v e^{-i \omega_1},r v)$ with the constrains $\lambda_5=0$, $\lambda_4=-\lambda_3$,  gives rise to 4 extra GBs and only to one neutral state. The simplicity of the analytical expressions for the three no vanishing masses ensures that the boundness constrain $\lambda_1>0$ in addition to $\lambda_3>0$ give positive masses. Thus in this case the Y points are superfluous. As in the previous cases, we expect the B points to be similar to the Y ones, because we choose our parameters centered in 1 in order not to have problems with unitarity. In conclusion, for this case only the G and R points are interesting. Moreover we expect that the most stringent bound is given by  the decay constrains and not by $TSU$: massless particles give a small contribution to the oblique parameters and due to the limited number of new particles (2 charged  degenerate scalars) $TSU$ should not deviate too much by the SM values. Indeed  in fig. \ref{fig.vvuomega1} it is shown that the oblique parameters at 3 $\sigma$ level  do not constrain at all the G points. For this reason we reported only the R points  in the upper panel of  fig. \ref{fig.vvuomega1}. By looking at  the plot $m_1-m_{ch_1}$ in fig. \ref{fig.vvuomega1} we see that with respect to the minima so far analyzed we have  much less points and that as expected there are cuts in correspondence of $m_Z$ and $m_W$.

In conclusion, the solutions for the alignment  $(v e^{i \omega_1}, v e^{-i \omega_1},r v)$ with $\lambda_5=0$, $\lambda_4=-\lambda_3$ are not easy to find, but the Higgs phenomenology does not completely rule out this vacuum configuration. We could introduce a weight to estimate how much a solution is stable or fine-tuned but this goes over the purposes of this work. We expect that this situation with 4 extra massless particles could be very problematic when considering the model dependent constraints \cite{ABMP:Constraining2}.

\mathversion{bold}
\subsubsection{ $(v e^{i \omega_1}, v e^{-i \omega_1},r v)$ case $ii)$}
\mathversion{normal}
\label{ressecnoCP3}

\begin{figure}[ht!]
  \centering
  \includegraphics[width=7cm]{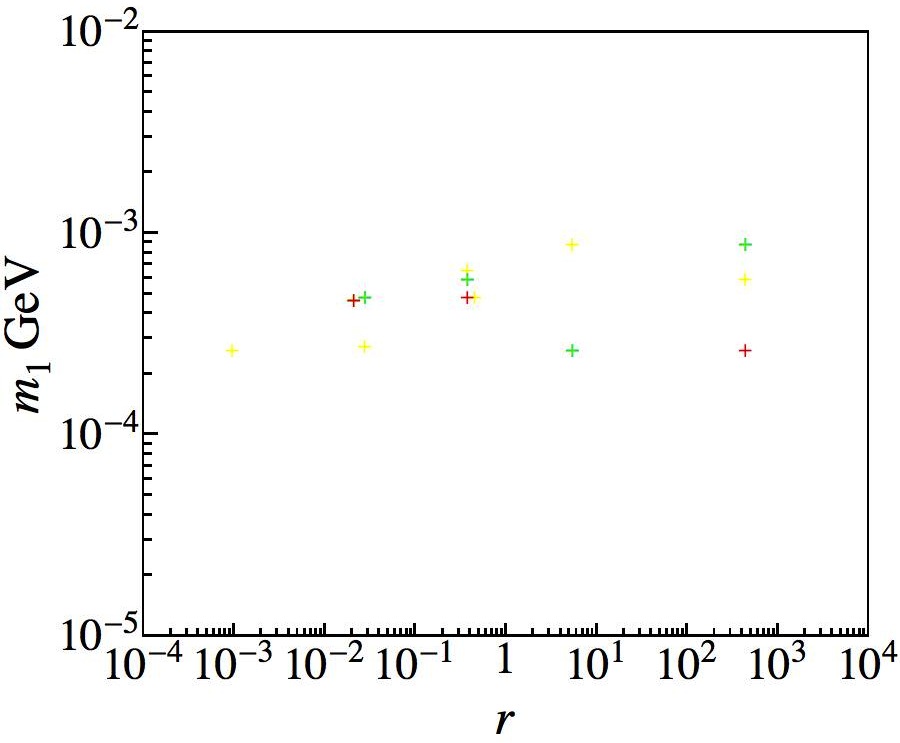}
   \includegraphics[width=7cm]{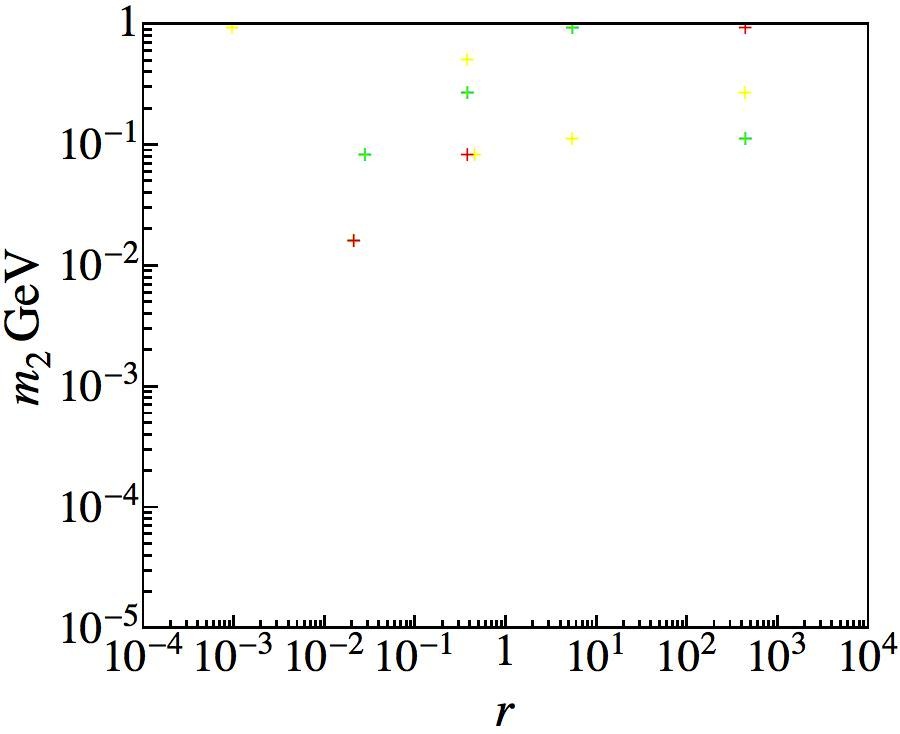}
    \caption{Alignment $(v e^{i \omega_1}, v e^{-i \omega_1},r v)$, case $ii)$: \it  the panels show $m_1$ (on the left) and $m_2$ (on the right) versus $r$. The number of points is small, but the interesting information is the order of magnitude of the masses.}
  \label{fig.vvuomega2}
\end{figure}

In the analytical discussion done in sec. \ref{seccomplecvevscaseii} we have seen that at least in the special limit $r\sim0$ ($r\sim1$ and $r>>1$) we expect the presence of one (two) very light particles. From all the numerical scans we performed we found out that solutions for the vacuum alignment $(v e^{i \omega_1}, v e^{-i \omega_1},r v)$ with the constraints of case $ii)$ are very difficult to be found. Moreover from fig. \ref{fig.vvuomega2} we see that for any value of $r$ the two lightest states  are always very light, thus confirming our rough analytical approximations. Indeed both $m_1$ and $m_2$ are lighter then we expected --especially $m_2$ for $r\sim 0$-- thus indicating that some cancellations have to occur to give all the masses greater then 0. This supports the difficulty to find solutions, difficulty that cannot to be ascribed to any constrain we imposed, because even in presence of 4 additional GBs as in sec. \ref{resseccomplexvev2} we found out a significant larger number of solutions.

The presence of a single R point in fig. \ref{fig.vvuomega2} is not statistically relevant, but more interesting is the order of magnitude of $m_{1,2}$: even in case $ii)$ we expect that the alignment  $(v e^{i \omega_1}, v e^{-i \omega_1},r v)$ may present serious problems once we add model dependent constraints \cite{ABMP:Constraining2}.

%
%
\section{Conclusions}
\label{sec:concl}

Flavour models based on non-Abelian discrete symmetries under which the SM scalar doublet (and its replicants) transforms non trivially are quite appealing for many reasons. First of all there are no new physics scales, since the flavour and the EW symmetries are simultaneously broken. Furthermore this kind of models are typically more minimal with respect to the ones in which the flavour scale is higher than the EW one: in particular the vacuum configuration is simpler and the number of parameters is lower. We then expect an high predictive power and clear phenomenological signatures in processes involving both fermions and scalars.

Due to the restricted number of parameters and the abundance of sensitive observables in these models, there are many constraints to analyze:  the most stringent ones arise by FCNC and LFV processes \cite{ABMP:Constraining2} but even Higgs phenomenology put several constraints on this class of models. The impact of the symmetry breaking in cosmology has been studied in \cite{Khlopov:book}.

In this paper we focussed on the $A_4$ discrete group, but this analysis can be safely generalized for any non-Abelian discrete symmetry. We consider three copies of the SM Higgs fields, that transform as a triplet of $A_4$. This setting has already been chosen in several papers \cite{MR:A4EWscale,LK:A4EWscale,MP:A4EWscale,Ma:A4EWscale} due to the simple vacuum alignment mechanism.

We have considered all the possible vacuum configurations allowed by the $A_4\times SM$ scalar potential. These configurations can account for both real and complex vevs. For all of them we have considered only model independent constraints, related to the Higgs-gauge boson Lagrangian, and postponing the model dependent analysis to an accompanying paper \cite{ABMP:Constraining2}. The first model independent constraint comes from the partial wave unitarity for the neutral two-particle amplitudes, which puts upper bounds on the scalar masses. Then we have explained how the light scalar mass region can be constrained considering the gauge boson decays. Moreover we  have seen how to put an upper bound on the lightest neutral state mass considering the Higgs decay channel $h\rightarrow W^+W^-$. Finally the most stringent bounds arise by the oblique parameters $TSU$.

We have shown that the Higgs-gauge boson model independent analysis can be used to study the parameter space of the difference vacuum configurations. Among the possible solutions which minimize the scalar potential, only one is ruled out due to the presence of tachyonic states. Furthermore, some other configurations may be obtained only by tuning  the potential parameters, giving rise to scalar spectrums characterized by very light or even massless particles. Finally, for the remaining ones, we find that they may share common features and this increases the difficulty in discriminating among them. Nevertheless, the model independent approach restricts in a non trivial way the parameter space.
In conclusion, we underline that more constraining results can be found considering specific realizations which adopt the different vacuum configurations: we present this analysis in \cite{ABMP:Constraining2}.

%
%

\section*{Note Added In Proof}
While completing this paper we received ref. \cite{MMP:A4ScalarPotential}, where the scalar potential with three copies of the SM Higgs doublet transforming as a triplet of $A_4$ is also studied. We stress the differences between this work an ours. Firstly, in \cite{MMP:A4ScalarPotential}, it is assumed that no new CP phases appear in the Higgs vevs, while we take this important possibility into account. Secondly, ref.\cite{MMP:A4ScalarPotential} discusses three interesting, but rather arbitrary vacua, where our analysis exhausts all possible vacua configurations. Lastly, a complete phenomenological study is missing in \cite{MMP:A4ScalarPotential}.

%
%

\section*{Aknowledgments}
We thank Ferruccio Feruglio for interesting comments and discussions. The work of RdAT and FB is part of the research program of the Dutch Foundation for Fundamental Research of Matter (FOM). The work of FB has also been partially supported by the Dutch National Organization for Scientific Research (NWO). RdAT acknowledges the hospitality of the University of Padova, where part of this research was completed. AP recognizes that this work has been partly supported by the European Commission under contract MRTN-CT- 2006-035505 and by the European Programme "Unification in the LHC Era", contract PITN-GA-2009-237920 (UNILHC).

%
%
\newpage
\mathversion{bold}
\appendixA{Appendix A: Analytical Formulae for $TSU$ Parameters}
\mathversion{normal}

In this Appendix we provide a sort of \emph{translator} from the papers \cite{PT:TSUparameters1,PT:TSUparameters2} to our notations and furnish the formulae we have used when different from their.

Reminding their notation we are in the case in which $n_d=3$ and $n_n, n_c=0$ so we do not have  the matrices  $\mathcal{T}$ and $\mathcal{R}$. Then we have
\bea
\mathcal{U}&\rightarrow & S\,\nn\\
\Re \mathcal{V}_{k i} &\rightarrow& U_{k i}\,,\nn\\
\Im \mathcal{V}_{k i}& \rightarrow &U_{k+3 i}\,,\nn\\
\omega_k&\rightarrow& f_k e^{i\omega_k}\,.
\eea
Moreover they put the GBs as first mass eigenstates while we put them as the last ones and contrary to them we use the standard definition for the photon.

We have rewritten they expression for
\beq
\label{deltaA}
\dfrac{A(I,J,Q)-A(I,J,0)}{Q}=
\begin{cases}
dA(I,J) & \mbox{ for $I\neq0$ and/or $J\neq 0$}\,,\\[2mm]
\dfrac{Q F(Q)}{Q}\sim \dfrac{1}{48 \pi^2} \log Q & \mbox{ for $I=J=0$ since $A(0,0,0)=0$}\,.
\end{cases}
\eeq
For the first row of \eq{deltaA} we have used
\beq
A(I,J,Q)\simeq A(I,J,0)+ Q\dfrac{\partial A(I,J,Q)}{\partial Q}\Bigg|_{Q=0}= A(I,J,0)+ Q\, dA(I,J) \,
\eeq
with
\beq
dA(I,J)=
\begin{cases}
\dfrac{1}{288 (I-J)^3 \pi ^2}\left[I^3+9 J I^2+6 (I-3 J) \log (I) I^2-9 J^2 I-J^3+6 (3 I-J) J^2 \log(J)\right]\\
\hspace{5cm}\,\mbox{for $I,J\neq 0,I\neq J$}\,,\\[2mm]
\dfrac{1}{288 \pi^2}(1+6 Log[I]) \qquad \qquad \mbox{for $J=0$}\,,\\[2mm]
\dfrac{1}{48 \pi^2}(1+\log[I]) \qquad \qquad \quad\;\, \mbox{for $I=J$} \,.
\end{cases}
\ee
The function $\bar{A}(I,J,Q)$ enters only in the loops in which a gauge boson and a scalar run, so we have always $J=Q$ when computing the quantity
\bea
\frac{\bar{A}(I,J,Q)-\bar{A}(I,J,0)}{Q}&=& \bar{dA}(I,J) \,.
\eea
As a result, for this function, it does not make sense considering the case $I=J=0$ being  $J=Q=m_V^2$ the  gauge boson mass.
We found
\beq
\bar{dA}(I,Q)=
\begin{cases}
\dfrac{1}{8 (I-Q)^3\pi ^2}\left[Q \left(-I^2+2 Q \log (I) I-2 Q \log (Q) I+Q^2\right)\right] & \mbox{for $I\neq Q,I\neq 0$}\,, \\[2mm]
\sim 0 \qquad\qquad \mbox{for $I=0$}\,,\\[2mm]
\sim 0 \qquad\qquad \mbox{for $I=Q$} \,.
\end{cases}
\eeq

\clearpage{\pagestyle{empty}\cleardoublepage}
%
%

\newpage
\bibliography{ConstrainingHiggs1v5.bbl}

\end{document}